\documentclass[fleqn,usenatbib]{mnras}
\usepackage{newtxtext,newtxmath, xcolor}

\usepackage[T1]{fontenc}

\DeclareRobustCommand{\VAN}[3]{#2}
\let\VANthebibliography\thebibliography
\def\thebibliography{\DeclareRobustCommand{\VAN}[3]{##3}\VANthebibliography}

\usepackage{graphicx}	
\usepackage{amsmath}	
\usepackage{amssymb}	
\usepackage[frozencache,cachedir=./mintedcache]{minted}     
\usepackage{hyperref}   

\newcommand{\parenth}[1]{\left(#1\right)}
\newcommand{\cosmo}[1]{#1_\text{cos}}

\newcommand{\spec}[1]{#1_\text{spec}}
\newcommand{\sort}[1]{#1_\text{sort}}
\newcommand{\phot}[1]{#1_\text{phot}}
\newcommand{\ctrl}[1]{#1_\text{ctrl}}
\newcommand{\reference}[1]{#1_\text{ref}}
\newcommand{\vir}[1]{#1_\text{vir}}
\newcommand{\SORT}{\textsc{sort}}
\newcommand{\tpcf}{2PCF}

\title[Galaxy Environment from SORT]{Galaxy Correlation Function and Local Density from Photometric Redshifts Using the Stochastic Order Redshift Technique (SORT)}

\author[James Kakos et al.]{
James Kakos,$^{1}$\thanks{E-mail: jkakos@ucsc.edu}
Joel R. Primack,$^{1}$
Aldo Rodr\'iguez-Puebla,$^{2}$
Nicolas Tejos,$^{3}$
\newauthor{ L. Y. Aaron Yung,$^{4}$
and Rachel S. Somerville$^{5}$}
\\
$^{1}$Physics Department, University of California, Santa Cruz, CA 95060, USA\\
$^{2}$Instituto de Astronom\'ia, Universidad Nacional Aut\'onoma de M\'exico, A. P. 70-264, 04510, CDMX, M\'exico\\
$^{3}$Instituto de F\'isica, Pontificia Universidad Cat\'olica de Valpara\'iso, Casilla 4059, Valpara\'iso, Chile\\
$^{4}$Astrophysics Science Division, NASA Goddard Space Flight Center, 8800 Greenbelt Rd, Greenbelt, MD 20771, USA\\
$^{5}$Center for Computational Astrophysics, Flatiron Institute, 162 5th Ave, New York, NY 10010, USA
}

\date{}

\pubyear{2021}

\begin{document}
\label{firstpage}
\pagerange{\pageref{firstpage}--\pageref{lastpage}}
\maketitle

\begin{abstract}
The stochastic order redshift technique (\SORT{}) is a simple, efficient, and robust method to improve cosmological redshift measurements. The method relies upon having a small (${\sim}10$ per cent) reference sample of high-quality redshifts. Within pencil-beam-like sub-volumes surrounding each galaxy, we use the precise $\mathrm{d}N/\mathrm{d}z$ distribution of the reference sample to recover new redshifts and assign them one-to-one to galaxies such that the original rank order of redshifts is preserved. Preserving the rank order is motivated by the fact that random variables drawn from Gaussian probability density functions with different means but equal standard deviations satisfy stochastic ordering. The process is repeated for sub-volumes surrounding each galaxy in the survey. This results in every galaxy with an uncertain redshift being assigned multiple `recovered' redshifts from which a new redshift estimate can be determined. An earlier paper applied \SORT{} to a mock Sloan Digital Sky Survey at $z \lesssim 0.2$ and accurately recovered the two-point correlation function on scales ${\gtrsim}4~h^{-1}$Mpc. In this paper, we test the performance of \SORT{} in surveys spanning the redshift range $0.75<z<2.25$. We used two mock surveys extracted from the Small MultiDark--Planck and Bolshoi--Planck $N$-body simulations with dark matter haloes that were populated by the Santa Cruz semi-analytic model. We find that \SORT{} is able to improve redshift estimates and recover distinctive large-scale features of the cosmic web. Further, it provides unbiased estimates of the redshift-space two-point correlation function $\xi(s)$ on scales ${\gtrsim}2.5~h^{-1}$Mpc, as well as local densities in regions of average or higher density. This may allow improved understanding of how galaxy properties relate to their local environments.
\end{abstract}




\section{Introduction}
In modern cosmology, the large-scale distribution of galaxies arises from the gravitational evolution and hierarchical clustering of primordial fluctuations. Large $\Lambda$CDM $N$-body simulations of cold dark matter and dark energy predict how these structures evolve. Such simulations show that many properties of dark matter haloes are correlated with the local density of the regions in which they form on scales of a few megaparsecs \citep[e.g.,][]{2017MNRAS.466.3834L}. But baryonic physics is complex, and we are still seeking to understand how galaxies form and evolve and how that is connected with the properties of their host dark matter haloes and the environments in which they reside \citep[e.g.,][and references therein]{2015ARA&A..53...51S,2018ARA&A..56..435W}. This can perhaps be clarified by comparing how halo properties and galaxy properties, such as stellar radius, depend on local density and small-scale clustering \citep[e.g.,][]{2021arXiv210105280B}.

We anticipate that this effort will be tremendously aided by the immense quantity of data that will flow from the giant LSST imaging survey at the Vera Rubin Observatory \citep{2019ApJ...873..111I} and the higher-resolution imaging surveys by the Euclid Space Telescope\footnote{\url{https://sci.esa.int/web/euclid}} and the Roman Space Telescope \citep{spergel2015widefield}. These surveys will provide multi-waveband photometric redshifts for billions of galaxies, of accuracy $\sigma_z /(1+z) \approx 0.02$ or better. Euclid will also measure grism redshifts of accuracy $\sigma_z / (1+z) \approx 10^{-3}$ for $\sim$30 million galaxies \citep{2021arXiv210801201S}. Meanwhile, the Dark Energy Spectroscopic Instrument \citep{2016arXiv161100036D} will measure redshifts of accuracy $\sigma_z/(1+z) \approx 10^{-4}$ for $\sim$10 million QSOs and $\sim$20 million galaxies, including $\sim$17 million emission-line galaxies in the redshift interval $0.6 < z < 1.6$. In the same regions of the sky where these accurate spectroscopic redshifts are being measured, the imaging surveys will produce more than an order of magnitude more photometric redshifts. It is therefore very important to develop methods that can make efficient use of the combination of a small fraction of spectroscopic redshifts and a much larger fraction of photometric or grism redshifts in order to measure the local environments and correlations of distant galaxies. The present paper discusses one such method.

The basic idea behind these methods is that galaxies cluster, especially on scales of a few megaparsecs. The idea of estimating redshifts using clustering was first developed by \citet{1979ApJ...227...30S,1987MNRAS.229..621P}; and \citet{1996ApJ...460...94L}. More recently, \citet{2013arXiv1303.4722M} proposed a method using a small set of reference galaxies with spectroscopic redshifts to estimate redshifts for a larger set of galaxies that are nearby on the sky to the reference galaxies. This was tested with simulations by \citet{2013MNRAS.431.3307S}, compared with spectroscopic redshifts by \citet{2015MNRAS.447.3500R}, used to reconstruct redshift distributions from measurement of the angular clustering of galaxies using a subset of spectroscopic redshifts by \citet{2016MNRAS.462.1683S}, and tested with simulations by \citet{2018MNRAS.474.3921S}. A related method was proposed by \citet{2017MNRAS.467.3576M}.

A method to estimate redshifts of galaxies with photometric redshifts using proximity to the cosmic web defined by a subset of galaxies with spectroscopic redshifts was proposed by \citet{2015MNRAS.454..463A}, who applied this PhotoWeb method to the SDSS out to redshift $z \approx 0.12$. \citet{2020A&A...636A..90S} applied this method to a larger sample of galaxies with spectroscopic redshifts from the SDSS and BOSS surveys out to redshift $z = 0.4$ to reconstruct the cosmic web using the DisPerSE algorithm \citep{2011MNRAS.414..350S}, and they used a convolutional neural network (CNN) trained with the SDSS and GAMA surveys to obtain photometric redshifts with mean absolute deviation $\sigma_{\rm MAD} \approx 0.01$ out to redshift $z \approx 0.3$ for bright galaxies with $r < 17.8$. They claimed that their version of the PhotoWeb method improved the accuracy of the redshifts by about a factor of two, to $\sigma/(1+z) \approx 0.004$.

The stochastic order redshift technique (\SORT{}; \citealp{2018MNRAS.473..366T}) is complementary to these approaches. It considers a patch on the sky where initially two kinds of galaxy redshift measurements exist: less accurate (e.g., photometric) and precise (spectroscopic). The galaxies with precise redshifts are used as a ‘reference sample,’ and it is of course expected that these correspond to a small fraction of the total number of galaxies. New `recovered` redshifts are sampled from the distribution of precise redshifts and matched one-to-one with the uncertain redshifts such that the rank order is preserved. This step is motivated by the fact that random variables drawn from Gaussian probability density functions (PDFs) with equal, arbitrarily-large standard deviations satisfy stochastic ordering. By construction, \SORT{} is non-parametric as it does not need to assume any functional form for either the distribution of redshifts or the relationship between spectroscopic and photometric redshifts. Thus, the power of \SORT{} relies on its simplicity and versatility. 

In this paper, we test how well the \SORT{} method can use photometric redshifts plus a smaller set of reference galaxies with spectroscopic redshifts to estimate the correlations of galaxies and the local densities of their environments out to high redshifts (here we focus on a redshift interval from 0.5 to 2.5). We test the \SORT{} method using mock galaxy surveys from backward light cones extracted from the Small MultiDark--Planck and Bolshoi--Planck cosmological $\Lambda$CDM simulations \citep{2016MNRAS.457.4340K,2016MNRAS.462..893R}. The dark matter haloes were populated with central and satellite galaxies using a current version of the Santa Cruz semi-analytic model (SAM), which has been shown to reproduce well the properties of observed galaxies out to high redshifts \citep[][and references therein]{2021MNRAS.502.4858S}. We show that \SORT{} is indeed robust and that it can provide unbiased measurement of the redshift-space two-point correlation function on scales ${\gtrsim}2.5~h^{-1}$Mpc while also recovering the local galaxy and mass density, especially in regions of higher than average density where most galaxies reside.

This paper is organized as follows. In Section~\ref{section:sort_method}, we briefly describe the method, while in Section~\ref{section:mock_survey} we describe the mock galaxy surveys used to study its performance. In Section~\ref{section:results}, we present the results of applying \SORT{} to a 2 square degree, mock wide-field light cone including galaxy redshifts, two-point correlation functions, and inferred three-dimensional densities of galaxy neighborhoods. In Section~\ref{section:discussion}, we provide a discussion regarding preservation of the redshift rank order, the effects of the \SORT{} parameters, limitations of the method, and potential future improvements. Section~\ref{section:summary} presents a summary and main conclusions. Appendix~\ref{appendix:candels_lc} provides the results of applying \SORT{} to a mock CANDELS light cone of area 0.2 square degrees. Appendix~\ref{appendix:satellite_coords} describes our method for assigning 3D coordinates to satellite galaxies in the Santa Cruz SAM. Appendix~\ref{appendix:sigma_ph_0.02} describes \SORT{} performance with larger photometric redshift uncertainties. Appendix~\ref{appendix:figure_code} describes the method we used to make vectorized figures with many points that are nevertheless of small file size. Appendix~\ref{appendix:additional_figures} provides several supplementary figures. All reported distances hereafter are comoving unless specified otherwise.

\section{The SORT Method} \label{section:sort_method}
Here we present a brief overview of the \SORT{} method. For a more complete discussion with illustrative figures, we refer the reader to \citet{2018MNRAS.473..366T}.

\subsection{General Idea}
Consider a set of $N$ galaxies comprised of a mixture of low-quality (referred to as photometric) and high-quality (referred to as spectroscopic) redshifts within a volume $V$. Assume that there are $N_\text{ph}$ galaxies with photometric redshifts and $N_\text{sp}$ galaxies with spectroscopic redshifts. When observing galaxies along some pencil-beam-like sub-volume, each of these subsets of galaxies will have a redshift probability distribution, $P_\text{ph}$ or $P_\text{sp}$, dependent upon their respective redshift uncertainties. In principle, both $P_\text{ph}$ and $P_\text{sp}$ can be considered representations of the same underlying true probability distribution with different levels of noise. Due to the greater expense of obtaining spectroscopic redshifts than photometric redshifts, the statistics for $P_\text{sp}$ are comparatively limited. However, if $N_\text{sp}$ is large enough to be statistically relevant to the total set of $N$ galaxies -- i.e. accurately traces the cosmic structure within the volume -- the higher quality of the spectroscopic redshifts will provide us with a higher resolution look at the true galaxy distribution. In this way, $P_\text{ph}$ can be considered a noisier version of $P_\text{sp}$.

We can leverage the relationship between $P_\text{ph}$ and $P_\text{sp}$ to try to improve the estimates of the photometric redshifts. To do this, we rely on stochastic ordering, which is defined as follows. Given two PDFs $P_A(x)$ and $P_B(x)$, the variable $X_A$ is stochastically less than $X_B$ if
\begin{equation} \label{eq:stochastic_ordering}
    P_A(X_A > x) \le P_B(X_B > x) \quad\forall x.
\end{equation}
To relate this to redshift estimates, consider two observed photometric redshifts $z_i^\text{ph}$ and $z_j^\text{ph}$ where $z_i^\text{ph} < z_j^\text{ph}$. We can think of each of these as being random variables sampled from Gaussian\footnote{The redshift PDFs are not required to be Gaussian, but this is used for simplicity.} PDFs centred on $z_i^\text{true}$ and $z_j^\text{true}$, respectively, with equal standard deviations determined by the measurement uncertainties.\footnote{Alternatively, one can think of the Gaussian PDFs as centred on $z^\text{ph}$ where the PDF corresponds to the probability of finding $z^\text{true}$ at a given $z$.} Even with potentially overlapping PDFs, $z_i^\text{ph}$ and $z_j^\text{ph}$ will satisfy stochastic ordering. Therefore, we can say the most likely scenario is that the underlying true redshifts satisfy $z_i^\text{true} \le z_j^\text{true}$. By extension, if we have $N_\text{ph}$ redshift estimates ordered such that $z_1^\text{ph} \le z_2^\text{ph} \le \ldots \le z_{N_\text{ph}}^\text{ph}$, we would also expect the true redshifts to most likely have the same rank ordering such that $z_1^\text{true} \le z_2^\text{true} \le \ldots \le z_{N_\text{ph}}^\text{true}$.

Of course, we do not have the `true' redshifts for galaxies, so we rely on high-quality spectroscopic redshifts instead. To apply the idea, we search in pencil-beam-like sub-volumes to determine $P_\text{sp}$ in that sub-volume. We then randomly sample $N_\text{ph}$ `recovered' redshifts, $z_i^\text{rec}$, from $P_\text{sp}$. Both the photometric redshifts and the recovered redshifts are rank ordered and matched one-to-one such that $z_i^\text{rec} \leftrightarrow z_i^\text{ph}$ for all $N_\text{ph}$ redshifts. In doing this, we take advantage of the higher resolution provided by $P_\text{sp}$ and simultaneously preserve the rank ordering.

We note that there may be cases where Eq.~\ref{eq:stochastic_ordering} does not hold true. However, we can expect that for state-of-the-art photometric redshift uncertainties, the PDFs will be well-behaved and obey stochastic ordering for the majority of cases. We also note that \SORT{} is a statistical model that should only be applied to sets of galaxies rather than individual measurements. Overall, \SORT{} can improve redshift estimates of a set, but it can also make a small fraction of individual measurements worse than the original photometric estimates. Indeed, in some cases, \SORT{} may return individual measurements that are inconsistent with a galaxy's original PDF (i.e. redshifts with errors larger than three times the photometric uncertainty).

\subsection{The SORT Algorithm} \label{section:algorithm}
For each galaxy $i$ in the sample with photometric redshifts, the following steps are taken (see also fig. 1 from \citealt{2018MNRAS.473..366T}):
\begin{enumerate}
    \item A circle with radius $R$ is drawn on the sky around the $i$th galaxy.
    \item Galaxies that fall within a cylinder defined by the radius $R$ and a redshift range $z_i\pm\Delta_z$ are selected and used for the remaining steps.
    \item From the selected galaxies, a check is made to ensure there are at least $N^\text{min}_\text{ref}$ galaxies with spectroscopic redshifts. If there are not at least $N^\text{min}_\text{ref}$ spectroscopic redshifts, the circle radius is incremented by $\delta R$ until the criterion is met or $R$ exceeds some $R_\text{max}$. If $R$ exceeds $R_\text{max}$, \SORT{} is considered to have failed and does not return any redshifts. The algorithm then moves to the next galaxy.
    \item A redshift histogram of the spectroscopic galaxies is made using a binning of $dz/3$. The histogram is then convolved with a Gaussian with $\sigma=dz$.\footnote{The motivation for this step is to have a smooth version of the discrete $\mathrm{d}N/\mathrm{d}z$ distribution associated with the reference sample.}
    \item For each of the photometric galaxies within the cylinder, a new recovered redshift is sampled from the histogram of spectroscopic redshifts.
    \item The selected galaxies' redshifts and the recovered redshifts are each rank ordered and matched one-to-one so each photometric galaxy is assigned a recovered redshift. \label{step:sorting}
\end{enumerate}
As this procedure is carried out for the remaining galaxies, every time a given galaxy is within the cylinder of one of its neighbors, it will gain another recovered redshift based on that selection. After the algorithm completes, each galaxy is assigned the median of all its recovered redshifts as its sorted redshift, $\sort{z}$. The values used for the algorithm parameters are discussed in Section~\ref{section:results}.

\section{Mock Galaxy Surveys} \label{section:mock_survey}

\subsection{Simulations and Backward Light Cones}
\begin{table*}
    \centering
    \caption{Comparison of the two mock light cones used. Each light cone was restricted to the redshift range $0.75<z<2.25$. Galaxies were selected from three complete redshift bins (as shown in Fig.~\ref{fig:completeness}). The light cones were extracted from different simulations, though the cosmological parameters are the same for both with the exception of $\sigma_8$.}
    \begin{tabular}{c|c|c|c|c|c|c|c|c|c|c}
    \hline
    Light Cone & Size (deg$^2$) & Galaxies & Completeness & Simulation & $\Omega_{\Lambda,0}$ & $\Omega_{\text{M},0}$ & $\Omega_{\text{B,0}}$ & $h$ & $n_s$ & $\sigma_8$\\
    \hline
    Wide Field & 2 & 1,058,366 & $H<27$ & Small MultiDark--Planck & 0.693 & 0.307 & 0.048 & 0.678 & 0.96 & 0.829\\
    CANDELS & 0.2 & 47,404 & $H<25.5$ & Bolshoi--Planck & 0.693 & 0.307 & 0.048 & 0.678 & 0.96 & 0.823\\
    \hline
    \end{tabular}
    \label{table:mock_comparison}
\end{table*}

We use mock galaxy surveys constructed by extracting dark matter haloes along backwards light cones from the Small MultiDark--Planck and Bolshoi--Planck $N$-body simulations \citep{2016MNRAS.457.4340K, 2016MNRAS.462..893R}. A brief summary of the light cones and their respective simulations is shown in Table~\ref{table:mock_comparison}. The dark matter haloes in the simulations were identified using \textsc{rockstar} \citep{2013ApJ...762..109B}. The backward light cones were constructed using the \texttt{lightcone} package\footnote{\url{https://bitbucket.org/pbehroozi/universemachine/src/master/}} released by \citet{2019MNRAS.488.3143B}, and further details are described in \citet{2021MNRAS.502.4858S} and \citet{2021ApJ...911..132Y}. 

The merger histories of the dark matter haloes were constructed using an algorithm based on the extended Press--Schechter formalism \citep{1999MNRAS.305....1S, 2008MNRAS.391..481S}. The formation and evolution of galaxies within these haloes was then modeled using the Santa Cruz SAM \citep{1999MNRAS.310.1087S, 2008MNRAS.391..481S,2015MNRAS.453.4337S}. \citet{2021MNRAS.502.4858S} presented a suite of light cones that was designed to represent the geometry and approximate areas of the five fields from the Cosmic Assembly Near-infrared Deep Extragalactic Legacy Survey (CANDELS)\footnote{\url{http://arcoiris.ucolick.org/candels/}}. They compared the mock survey predictions with the CANDELS observed counts, stellar mass functions, rest-frame luminosity functions from $0.1 \lesssim z \lesssim 2$, and found generally good agreement. Yung et al. (in preparation) present a suite of 2 square degree mock light cones, that have been populated with galaxies using the same approach. In this work, we make use of one of the 2 square degree mock light cones and one of the mock CANDELS catalogues with field geometry similar to the COSMOS field, covering an area on the sky of $17\times41$ square arcmin in right ascension and declination.

The Santa Cruz SAM does not make use of the $N$-body positions and velocities for dark matter haloes once they become ``sub-haloes'' (or satellites) within a larger halo. Instead, it estimates the galactocentric radius of each satellite from the centre of the halo and its decay due to dynamical friction using an analytic model (see \citet{2008MNRAS.391..481S} and \citet{2021MNRAS.502.4858S}). As a result, in order to compute separate redshifts for the satellites, the 3D positions and velocities must be assigned in post-processing. For details on our method for assigning these properties to the satellite galaxies, see Appendix~\ref{appendix:satellite_coords}.

The full catalogues span the range $0<z<10$, though this work uses only galaxies in the range $0.75<z<2.25$ based on their mock observed redshifts. The lower redshift limit was imposed to ensure the light cones had large enough cross-sectional areas to measure the two-point correlation function on scales of ${\gtrsim}3~h^{-1}$Mpc. Mock observed redshifts were calculated using
\begin{equation} \label{eq:z_obs}
    z_\text{obs} = z_\text{los} + \delta_z (1+z_\text{los})
\end{equation}
where $z_\text{los}$ is the redshift that includes distortions from peculiar velocities along the line of sight and $\delta_z$ is a random sample from a Gaussian centred at zero with standard deviation $\sigma_z$ (either photometric or spectroscopic). Note that we do not model catastrophic failures in the photometric sample as we do not expect them to have a significant effect on the net result of the \SORT{} method. 

Apparent magnitudes are provided in the mock light cones. For this work, we use $H$-band magnitudes given by the `wfc3f160w\_dust' output of the Santa Cruz SAM. The observed-frame IR luminosities are calculated based on the star formation histories predicted by the Santa Cruz SAM and stellar population synthesis models of \citet{2003MNRAS.344.1000B}. Dust attenuation is modeled using a standard `slab' model as described in \citet{2012MNRAS.423.1992S}. For more details, we refer the reader to \citet{2021MNRAS.502.4858S}.

\begin{figure}
    \centering
    \includegraphics[width=\linewidth]{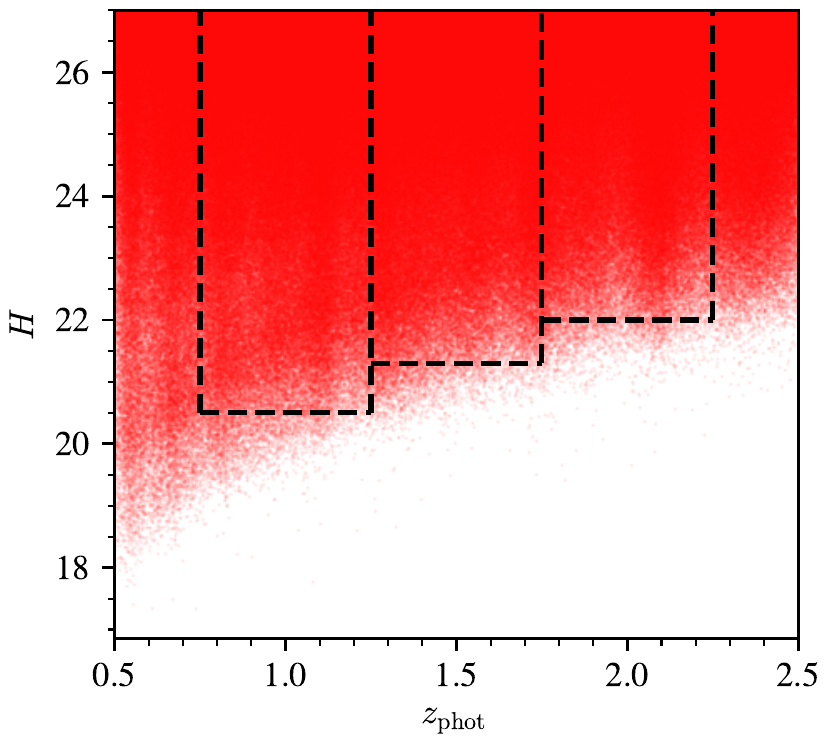}
    \caption{Galaxies were selected from the three volume-complete regions defined by the dashed lines. Spectroscopic redshifts were assigned within each region randomly to 10 per cent of the galaxies in that region. See Appendix~\ref{appendix:figure_code} for a note on how this figure (and others in this paper) were made with a combination of vectorized and rasterized elements that maintain clarity while reducing file sizes.}
    \label{fig:completeness}
\end{figure}

We adopt a completeness of $H<25.5$ for the mock CANDELS light cone, which is a rough limit to which we expect CANDELS photometric redshifts to be accurate. We expect future surveys to improve this and thus adopt $H<27$ for the wide-field light cone. Galaxies were selected from three volume-complete regions within the light cone, as shown in Fig.~\ref{fig:completeness}. In each of the three regions, 10 per cent of galaxies were randomly chosen to have mock spectroscopic redshifts while the remaining 90 per cent were given mock photometric redshifts. After preparing the mock catalogues, the 2 square degree wide-field light cone had 1,058,366 galaxies (${\sim}147$ galaxies per square arcmin) and the mock CANDELS light cone had 47,404 galaxies (${\sim}$68 galaxies per square arcmin). All results in the main text of this paper are drawn from the wide-field light cone as this provides better overall statistics. Parallel results for the mock CANDELS light cone are shown in Appendix~\ref{appendix:candels_lc} and are more representative of present galaxy surveys.

\subsection{Redshift Types}
Here we define several different redshift types that will be discussed in our results:
\begin{description}
    \item $\cosmo{z}$ : These are redshifts that are purely cosmological and include neither redshift-space distortions from line-of-sight peculiar velocities nor measurement uncertainty.
    \item $z_\text{los}$ : These are the `true' redshifts that include line-of-sight peculiar velocities.
    \item $\spec{z}$ : These are simulated spectroscopic redshifts that include a small measurement uncertainty according to Eq.~\ref{eq:z_obs}.
    \item $\reference{z}$ : These are the reference sample redshifts. They comprise a relatively small fraction of the total number of redshifts and have spectroscopic quality.
    \item $\phot{z}$ : These are simulated photometric redshifts. They are generated the same way as $\spec{z}$ but with larger uncertainties.
    \item $\sort{z}$ : These are the results of running the \SORT{} method.
    \item $\ctrl{z}$ : These are the results of the controlled \SORT{} algorithm that excludes rank ordering (see Section~\ref{section:control_test} for details).
\end{description}

\section{Results} \label{section:results}
Most of the results in this paper were obtained assuming a spectroscopic redshift fraction of 10 per cent, although we also explored larger and smaller spectroscopic fractions (see Fig.~\ref{fig:dz_hist_fg}). The spectroscopic and photometric uncertainties used were $\sigma^\text{sp}_z/(1+z)=0.0001$ and $\sigma^\text{ph}_z/(1+z)=0.01$, respectively, but we also provide results for $\sigma^\text{ph}_z/(1+z)=0.02$ in Figure~\ref{fig:2pcf_3styles} and Appendix~\ref{appendix:sigma_ph_0.02}. The minimum required number of reference galaxies for each sub-volume was set to $N_\text{ref}^\text{min}=4$. \SORT{} is effective with this value as low as $N_\text{ref}^\text{min}=2$, but we found increasing to 4 provided a better overall estimate of the two-point correlation function. The initial search radius was set to $R=0.01^\circ$ and the redshift bin width was set to $dz=0.0003$. These correspond to length scales of around 0.3--0.7~$h^{-1}$Mpc for $R$ and 0.3--0.6~$h^{-1}$Mpc for $dz$ in the range $0.75<z<2.25$. These values were chosen to be able to capture relevant scales of the cosmic web. The search radius increment was set to $\delta R=0.1R$ with a maximum possible radius of $R_\text{max}=0.1^\circ$. If the $N_\text{ref}^\text{min}$ criterion was not met within $R\le R_\text{max}$ for a given galaxy, that galaxy was removed from the results.\footnote{In the data presented, no such galaxies were removed.} The search depth was limited to $z_i \pm \Delta_z$ with $\Delta_z = 2.5\sigma^\text{ph}_z$. This depth was chosen to be large enough to capture nearly all photometric redshifts and their true environments within the same sub-volume. See Section~\ref{section:selection_volume} for details on these parameters.

Our primary comparison for the results of \SORT{} is to $\spec{z}$, as spectroscopic redshifts represent our best estimates of galaxy redshifts and \SORT{} uses these to trace the cosmic web. However, in some cases, we also show results of $\cosmo{z}$ despite these redshifts not being directly observable due to redshift-space distortions. These results are shown for comparison as they represent the true underlying distribution of galaxies. Fig.~\ref{fig:square_field} and Fig.~\ref{fig:square_field_alt} show $\cosmo{z}$ to illustrate the effects of redshift-space distortions and the alignment of reference galaxies with the true cosmic web. Density estimates of $\sort{z}$ are compared to those of $\cosmo{z}$ in Fig.~\ref{fig:all_density_4Mpc} as \citet{2017MNRAS.466.3834L} showed that many halo properties correlate with local densities using the true $N$-body positions of haloes (which are replicated by using $\cosmo{z}$, not $\spec{z}$).

\subsection{Improving Redshift Estimates}

\begin{figure}
    \centering
    \includegraphics[width=\linewidth]{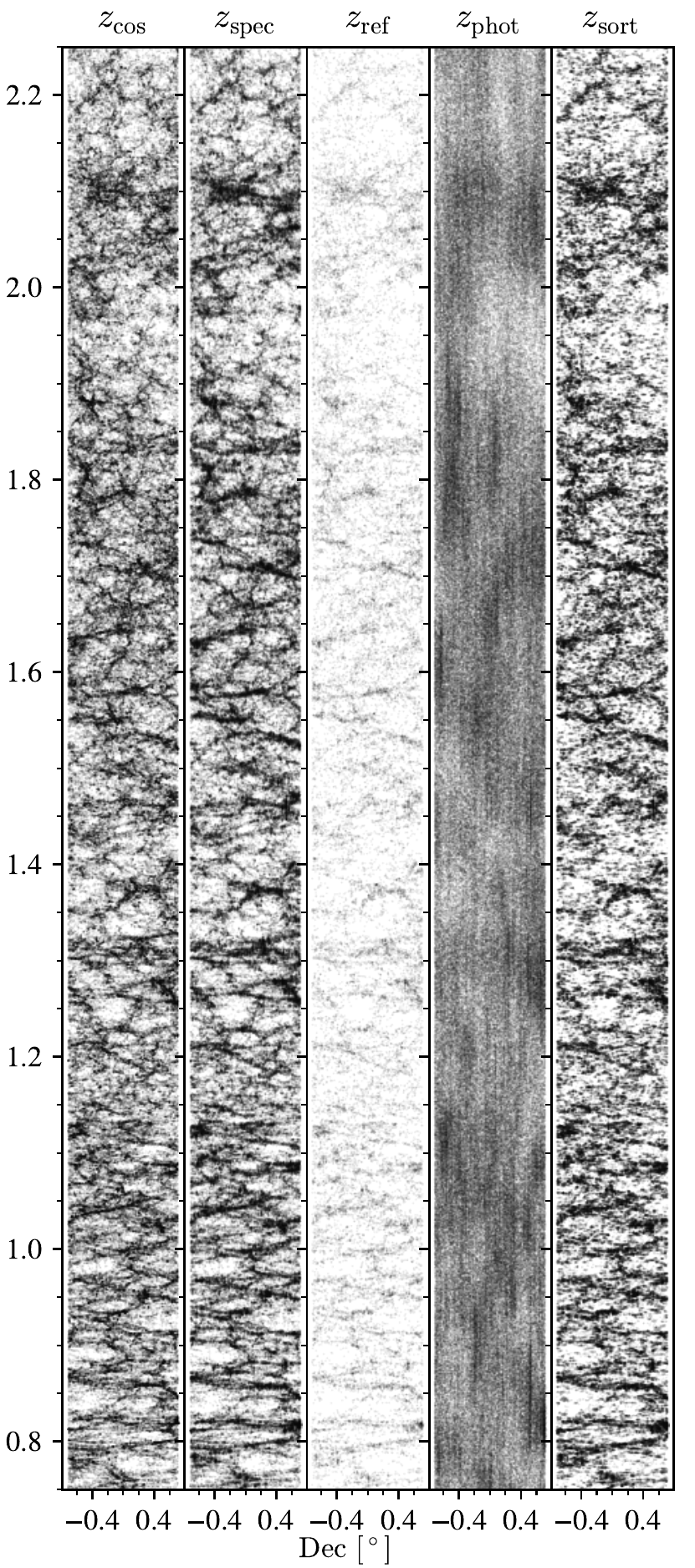}
    \caption{Scatter plot of the projected two-dimensional distribution of galaxies for $\cosmo{z}$, $\spec{z}$, $\reference{z}$, $\phot{z}$, and $\sort{z}$. Each panel shows a $0.5^\circ$ slice in right ascension and the full declination of the light cone. The middle panel, $\reference{z}$, corresponds to 10 per cent of the total galaxies, and the remaining panels show the 90 per cent non-reference galaxies. The large-scale features of the cosmic web are much more identifiable with $\sort{z}$ than $\phot{z}$. However, \SORT{}'s tendency to group galaxies closely together means that it struggles to recover low-density regions. Note that the horizontal cuts slightly visible in the $\phot{z}$ panel are a result of the completeness condition shown in Fig.~\ref{fig:completeness}.}
    \label{fig:big_field}
\end{figure}

\begin{figure*}
    \centering
    \includegraphics[width=\linewidth]{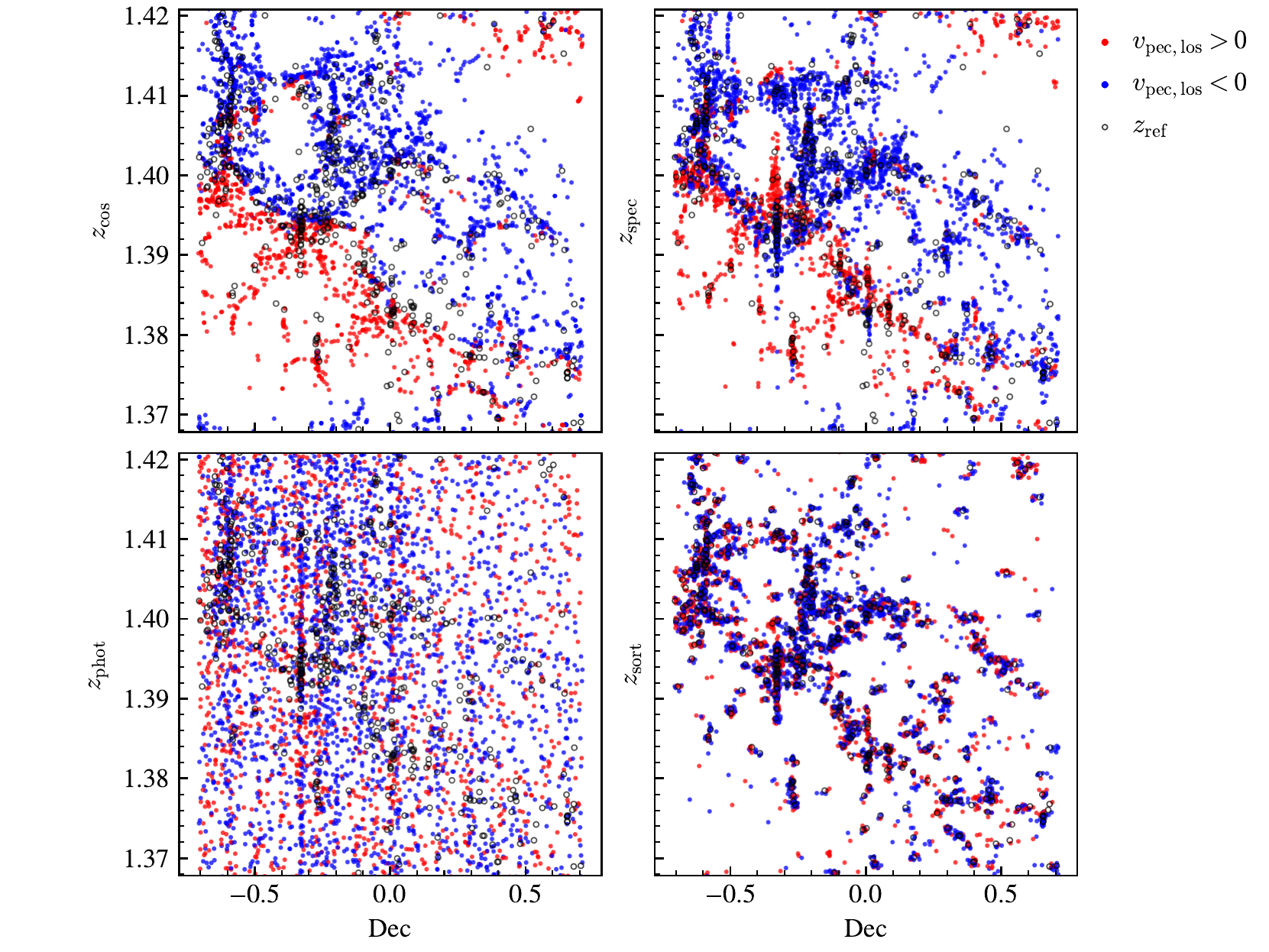}
    \caption{Right ascension slices (thickness 0.1$^\circ$) of galaxy distributions using different redshifts in a roughly $75\times75~h^{-1}$Mpc region of space. The red and blue colouring denotes the direction of the peculiar velocity along the line of sight (red is positive and blue is negative). The black rings with empty centres are reference galaxies. This region is dominated by a high-density ring of galaxies that surrounds a void in the upper left quadrant. We can see that the accurate tracing of this ring by the reference sample allows \SORT{} to recreate it while also preserving the void in the centre. We expect that such voids surrounded by a sufficiently high density of galaxies should largely be preserved in $\sort{z}$. Reference galaxies are rarely found in voids, but may be shifted into them by redshift-space distortions in cases where dense clusters are positioned along the line of sight to the void. As such, \SORT{} will primarily place galaxies around the voids where the reference galaxies reside. On the other hand, with photometric redshifts, the large uncertainties in the measurements of galaxies surrounding the void smooth out the region, obscuring the underlying structure as shown in the bottom left panel.}
    \label{fig:square_field}
\end{figure*}

A general look at how well \SORT{} is able to improve redshift estimates can be seen in Fig.~\ref{fig:big_field}. Each panel shows a different redshift type plotted against the full declination of the light cone. The middle panel shows the reference sample, $\reference{z}$, which is comprised of 10 per cent of the spectroscopic sample and is assumed to be known when \SORT{} is applied. This is the structural outline that \SORT{} uses to reassign redshifts. In the $\phot{z}$ panel, the cosmic structure is almost entirely smoothed out. Even with an optimistic photometric uncertainty of $\sigma^\text{ph}_z/(1+z)=0.01$, one can only get a very rough sense of high- or low-density regions. The $\sort{z}$ panel shows a significant improvement on $\phot{z}$. We see more accurate clustering of galaxies, as well as signs of filamentary structure and voids. We note that \SORT{}'s reconstruction of low-density regions is not particularly good. This is primarily due to \SORT{}'s tendency to place galaxies near other galaxies. Lower-density regions will be populated with few galaxies, and only a small fraction of those will be reference galaxies.

A more zoomed-in view of the different redshift types can be seen in Fig.~\ref{fig:square_field}. Each panel shows  a square region of space, roughly $75\times75~h^{-1}$Mpc. The red and blue colouring represents the direction of the peculiar velocities along the line of sight; red points have positive velocities and blue points have negative velocities. When comparing $\cosmo{z}$ to $\spec{z}$, we see that galaxies in denser regions become spread out vertically. The severity of these distortions will directly impact \SORT{}'s ability to reconstruct the cosmic web. Redshift-space distortions in the reference sample will inherently affect how \SORT{} assigns redshifts. For example, there is a dense cluster of galaxies in Fig.~\ref{fig:square_field} in front of a void. The redshift-space distortions cause a number of galaxies, including some reference galaxies, to be shifted into the void. This results in \SORT{} placing galaxies in the void where they otherwise should not be placed.

We notice also how \SORT{} clusters galaxies tightly to the reference sample. In the lowest density environments, there are cases where galaxies build up around one or two reference galaxies -- e.g., around (0.5, 1.46) in the $\sort{z}$ panel. Galaxies are pulled along the line of sight to a nearby reference galaxy, leading to horizontal structures in a plane perpendicular to the line of sight. This is most prominent in low density environments because \SORT{} has to increase its search radius to find reference galaxies. This allows galaxies at a wider range of angular separations to be placed at roughly the same redshift. Overall, though, we see that \SORT{} does a fairly good job at reconstructing the main features in this region of space, especially compared to the photometric redshifts. In the $\phot{z}$ panel on the lower left, any sign of the main features in this region is almost completely lost.

\begin{figure*}
    \centering
    \includegraphics[width=\linewidth]{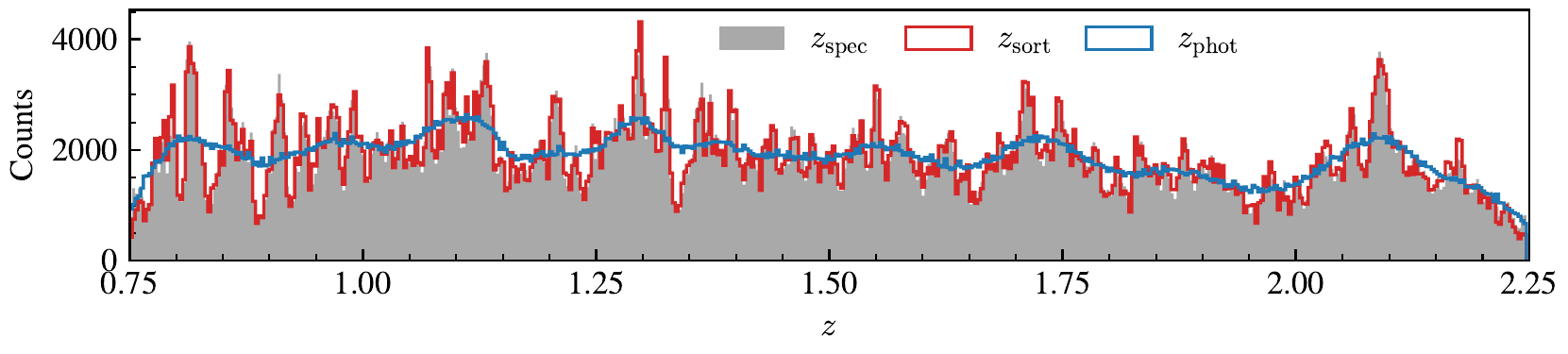}
    \caption{Redshift distributions for $\spec{z}$ (grey), $\sort{z}$ (red), and $\phot{z}$ (blue) with arbitrary binning of 0.003. The large uncertainty of the photometric redshifts blurs out the structure of the distribution, which becomes more or less flat over the entire range. The distribution produced by \SORT{} much more closely follows the distribution obtained with spectroscopic measurements. This is by design, as \SORT{} samples new redshifts based on the distribution of the spectroscopic-quality reference sample within each sub-volume.}
    \label{fig:z_dist1d}
\end{figure*}

This is further shown when looking at the recovery of the spectroscopic $\mathrm{d}N/\mathrm{d}z$ distribution. The one-dimensional redshift distributions are shown in Fig.~\ref{fig:z_dist1d} for $\spec{z}$ (grey region), $\sort{z}$ (red), and $\phot{z}$ (blue). The peaks and valleys are smoothed out in the photometric distribution while $\sort{z}$ shows significant improvement in ability to outline large-scale structure along the line of sight. By construction, $\sort{z}$ is meant to follow the $\mathrm{d}N/\mathrm{d}z$ distribution of the spectroscopic reference sample, and that is what we observe here. Some of the discrepancy is a result of \SORT{} pulling galaxies from low-density regions, where reference galaxies are scarce, to high-density regions.

\begin{figure}
    \centering
    \includegraphics[width=\linewidth]{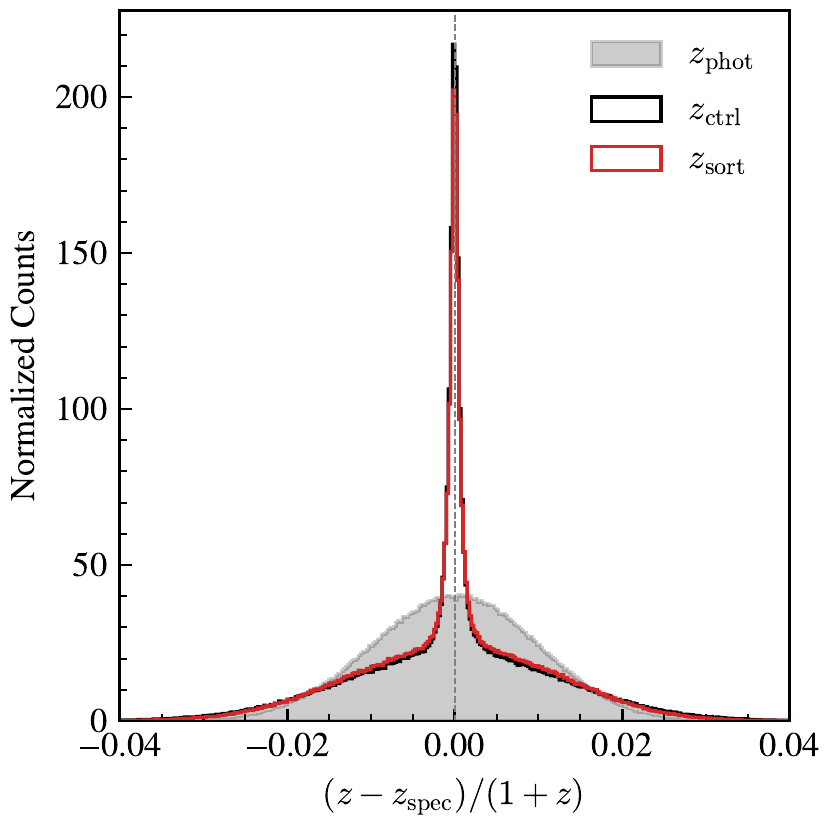}
    \caption{Normalized distribution of $\Delta z$ (excluding the spectroscopic sample) for $\sort{z}$, $\phot{z}$, and $\ctrl{z}$ (see Section~\ref{section:control_test} for details on $\ctrl{z}$.). The photometric distribution essentially recovers the Gaussian used to create the photometric sample. \SORT{} is able to produce a tall peak surrounding $\Delta z=0$ where a significant fraction of redshifts have been improved. The overall standard deviation of $\Delta\sort{z}$ is comparable to $\Delta\phot{z}$ as shown by the broader base of the distribution.}
    \label{fig:dz_hist}
\end{figure}

Fig.~\ref{fig:dz_hist} shows the error $\Delta z/(1+z)$ with respect to $\spec{z}$. The grey shaded region shows the error distribution for $\phot{z}$. Since the photometric redshifts were generated using a Gaussian distribution, the recovered distribution is Gaussian with a standard deviation of ${\sim}0.01(1+z)$. In red, the results of \SORT{} show a significant fraction of redshifts that have been improved. Overall, $\Delta\sort{z}$ and $\Delta\phot{z}$ share a similar standard deviation; however, the large peak shows that $\sort{z}$ provides much more information than $\phot{z}$. This is shown clearly in both Fig.~\ref{fig:big_field} and Fig.~\ref{fig:z_dist1d} as $\sort{z}$ is able to more accurately outline large-scale structure that is washed out by $\phot{z}$.

\begin{figure*}
    \centering
    \includegraphics[width=\linewidth]{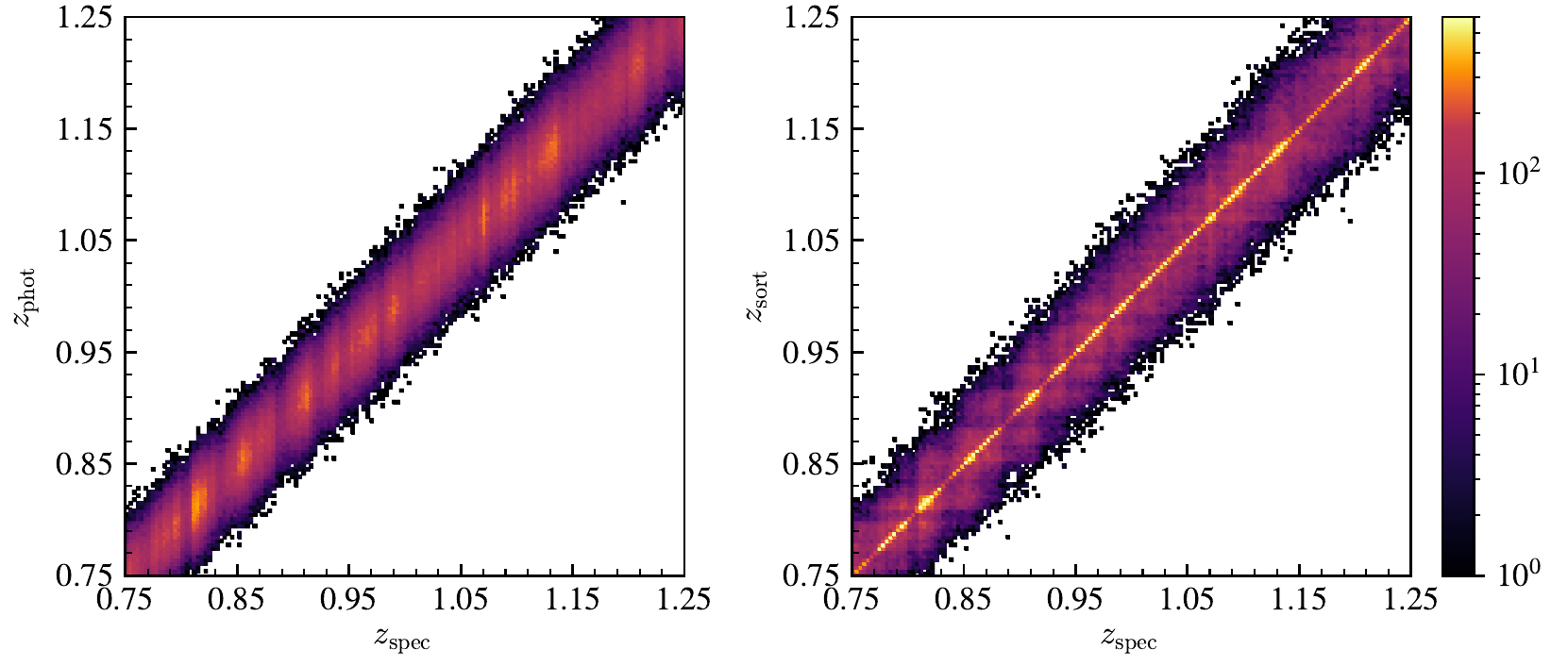}
    \caption{Two-dimensional redshift histograms for $\phot{z}$ and $\sort{z}$ relative to $\spec{z}$ with binning of 0.003. The color bar represents the total number of counts in each bin. We observe significant improvement in redshift estimates by $\sort{z}$ compared to $\phot{z}$. There are much higher counts along the line of equality for $\sort{z}$, and this effect is consistent across the entire redshift range of the light cone. All redshift bins can be seen in Fig.~\ref{fig:individual_z_dist2d}.}
    \label{fig:z_dist2d}
\end{figure*}

A direct comparison of redshifts can be seen in Fig.~\ref{fig:z_dist2d}. The left and right panels show the two-dimensional histograms of $\phot{z}$ and $\sort{z}$ compared to $\spec{z}$. We continue to see improvement in redshift estimates after applying \SORT{}. The large peak shown in Fig.~\ref{fig:dz_hist} is now represented by a bright, narrow band of redshifts along the line of equality where errors are small. This improvement is seen in all redshift bins, which are shown in Fig.~\ref{fig:individual_z_dist2d}.

\subsection{Recovering The Two-Point Correlation Function}
The two-point correlation function (\tpcf{}) is a relatively simple metric that provides information about the three-dimensional spatial clustering of galaxies. The large uncertainties associated with photometric redshifts lead to smoothing of spatial clustering and a highly biased estimate of the 3D \tpcf{} on relevant scales. As a result, using only photometric redshifts, one typically calculates the 2D angular \tpcf{}. Here we test \SORT{}'s ability to recover the 3D \tpcf{}. We note, however, that this test is somewhat conservative because redshift distortions and \SORT{} only affect positioning along the line of sight. Angular correlations do not deviate from their true values.

Estimates of the \tpcf{} were calculated using various redshift types as a function of redshift-space distance $s$. Both $\cosmo{\xi}(s)$ and $\spec{\xi}(s)$ assume 100 per cent of the galaxies have a known cosmological or spectroscopic redshift.\footnote{It is not expected that \SORT{} should recover $\cosmo{\xi}(s)$ since $\reference{z}$ traces $\spec{z}$, not $\cosmo{z}$. These results are shown for the sake of comparison.} $\reference{\xi}(s)$ uses only the reference sample -- i.e. only 10 per cent of galaxies with spectroscopic redshifts. For $\phot{\xi}(s)$, $\sort{\xi}(s)$, and $\ctrl{\xi}(s)$ (see Section~\ref{section:control_test} for details on $\ctrl{\xi}(s)$), \tpcf{}s were calculated using their respective 90 per cent non-reference sample redshifts plus the 10 per cent spectroscopic-quality reference sample. The \tpcf{}s were calculated using \textsc{corrfunc} \citep{2020MNRAS.491.3022S} from scales of ${\sim}1~h^{-1}$Mpc to ${\sim}18\text{--}30~h^{-1}$Mpc (larger scales are calculated in higher redshift bins; see Fig.~\ref{fig:2pcf_full_range} for \tpcf{}s in all redshift bins).

Using only photometric redshifts leads to a very poor estimate of the \tpcf{}. This is not surprising due to the large uncertainty associated with $\phot{z}$. On the other hand, we observe that \SORT{} accurately recovers the spectroscopic \tpcf{} on scales of $s\gtrsim2.5~h^{-1}$Mpc. At the smallest scales, however, \SORT{} overestimates the \tpcf{}. This result stems from the design of the \SORT{} algorithm -- namely, galaxies will be placed near other galaxies, resulting in high clustering on smaller scales.\footnote{In the previous \SORT{} paper \citep{2018MNRAS.473..366T}, the \tpcf{} was \textit{underestimated} on small scales. This difference stems from the fact that the previous paper did not include satellite galaxies while this one does.} The lower limit to which \SORT{} is accurate will depend on the choice of binning the method uses. As our chosen bin width corresponds roughly to $1~h^{-1}$Mpc, we can only expect to be reasonably accurate at scales larger than this.

\begin{figure}
    \centering
    \includegraphics[width=\linewidth]{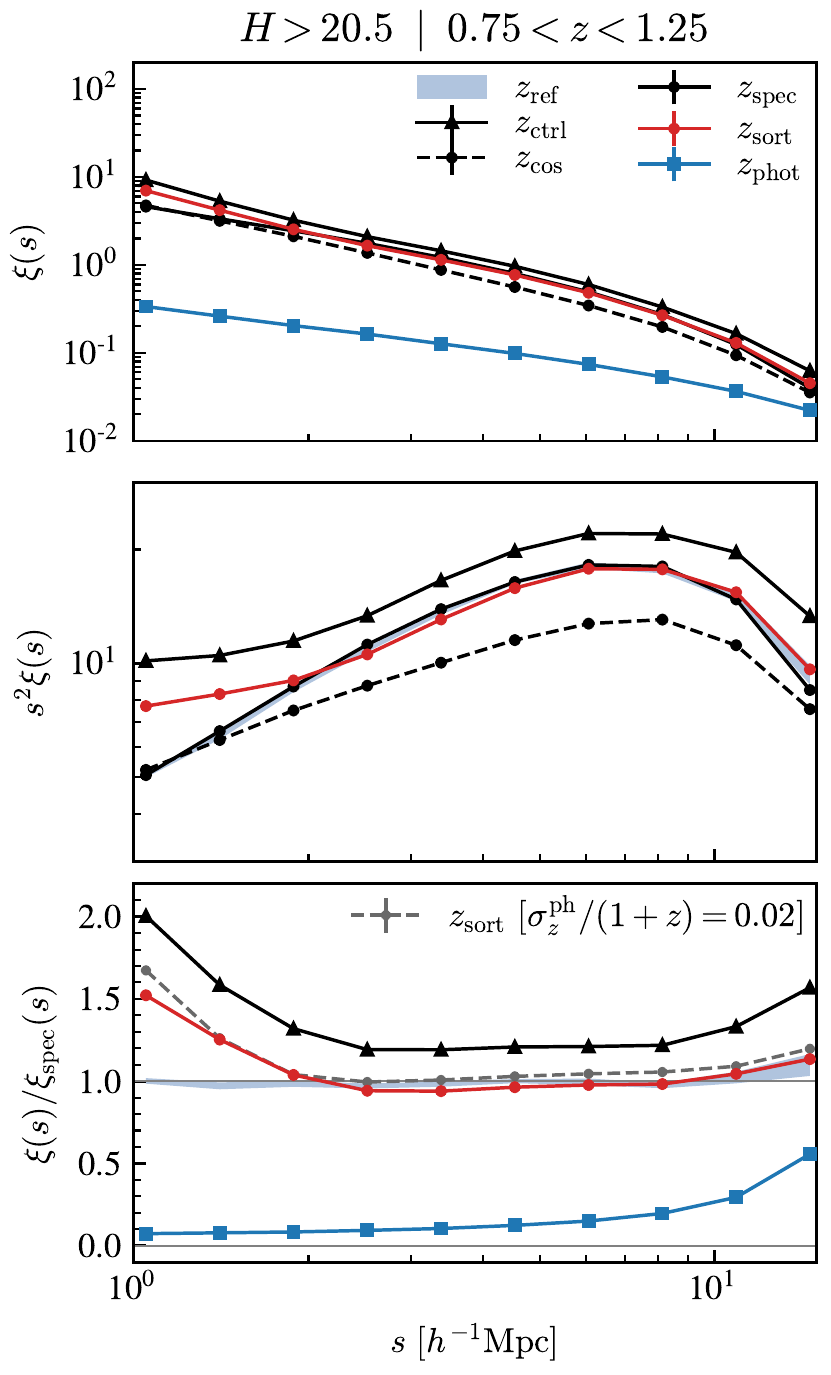}
    \caption{Two-point correlation functions (\tpcf{}s) of various redshift types as a function of redshift-space distance $s$ shown in three different ways. The results show the mean value of the \tpcf{}s along with 1$\sigma$ error bars after 10 bootstrap samples. Note that the error bars are too small to be seen. We observe that $\phot{\xi}(s)$ is a poor estimate of the \tpcf{} at all shown length scales, and $\ctrl{z}$ consistently overestimates the \tpcf{} while $\sort{\xi}(s)$ is accurate (relative to $\spec{\xi}(s)$) for $s\gtrsim2.5~h^{-1}$Mpc. See Section~\ref{section:control_test} for details on $\ctrl{z}$ (shown as black triangles) and Appendix~\ref{appendix:sigma_ph_0.02} for details on $\sort{\xi}(s)$ using $\sigma_z^\text{ph}/(1+z)=0.02$ (shown as the grey dashed line in the bottom panel).}
    \label{fig:2pcf_3styles}
\end{figure}

\subsection{Estimating Local Densities} \label{section:density}
Densities were calculated by searching for neighbors within cylindrical sub-volumes surrounding each galaxy. The total length of each cylinder was set to 4~$h^{-1}$Mpc. Photometric redshift uncertainties correspond to scales of $\sigma_z^\text{ph}=0.01(1+z)\approx30~h^{-1}$Mpc for $0.75<z<2.25$. This is much larger than the scale of the cylinder and thus photometric density estimates will be particularly poor. Nevertheless, we use this cylinder length to test the limits of \SORT{}. We also provide a sample of density estimates using a much longer cylinder defined by $l = 2\frac{\Delta v}{c} (1+z)$ with $\Delta v = 1000~\text{km~s}^{-1}$ in Fig.~\ref{fig:all_density_1000kms}.

The radius of the cylinder is initially set to $r=0.02^\circ$. If there are not at least $n$ galaxies within the cylinder, the radius is incremented by $\delta r=0.001^\circ$ until that condition is met or $r$ reaches $r_\text{max}=0.04^\circ$. This method was chosen to make the calculation adaptive. The range of densities across the entire light cone is large, and having an adaptive aperture allows for probing different scales. The radius can start small to inspect high-density regions and expand in low-density regions to estimate an average density where there may otherwise be only one or two galaxies in the cylinder. The values for $r$, $\delta r$, and $r_\text{max}$ were arbitrarily chosen to be similar to the parameters used by \SORT{}. Likewise, the minimum threshold of neighbors was arbitrarily chosen to be $n=5$. In principle, these parameters are all adjustable depending on how much ones wishes to constrain the densities. The results of \SORT{}'s estimations of local densities compared to photometric estimates are generally not sensitive to the choice in these parameters, though.

\begin{figure*}
    \centering
    \includegraphics[width=6in]{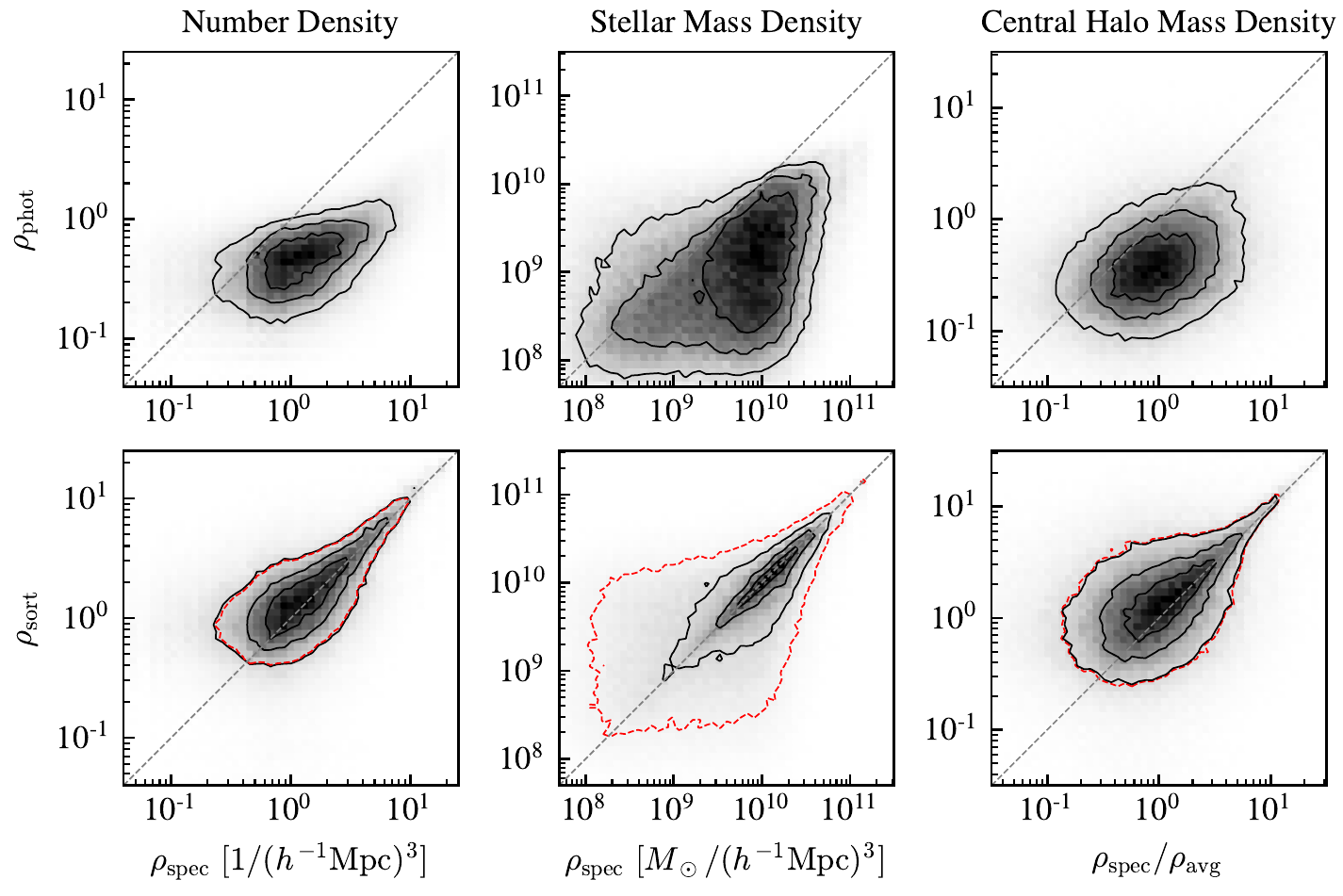}
    \includegraphics[width=6in]{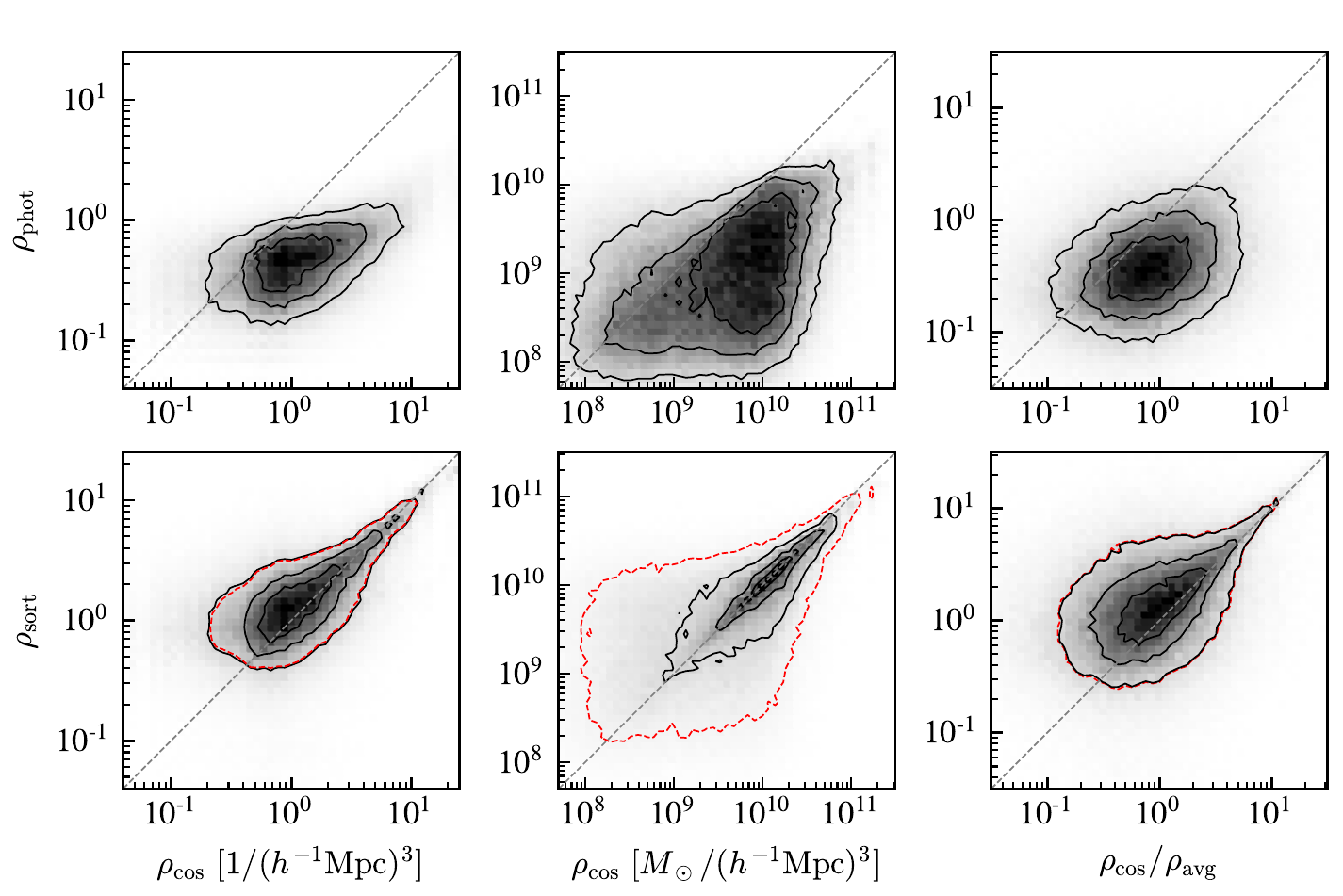}
    \caption{Two-dimensional density histograms for $\phot{\rho}$ and $\sort{\rho}$ in the range $0.75<z<1.25$. The left panels show number densities, the middle panels show stellar mass densities, and the right panels show halo mass densities using only central galaxies. The top six panels compare densities to estimates using $\spec{z}$, and the bottom six panels compare densities to estimates using $\cosmo{z}$. The solid contours represent limits of 25, 50, and 75 per cent of the maximum bin value in each subplot. The dashed contour (red) is set at a limit equal to the minimum contour level in the corresponding $\phot{\rho}$ subplot. Densities were calculated with a fixed cylinder length of 4~$h^{-1}$Mpc and a radius starting at ${\sim}1~h^{-1}$Mpc and expanding up to ${\sim}2~h^{-1}$Mpc as needed to encompass at least five galaxies. As expected, the photometric densities estimates are all poor as the cylinder length scales are much smaller than the typical photometric redshift error. \SORT{} densities show significant improvement in regions with average or higher density. Overall scatter is similar when comparing $\sort{\rho}$ to $\phot{\rho}$, but $\sort{\rho}$ displays much better alignment with the line of equality and a more peaked distribution surrounding it. See~Section~\ref{section:density} for details on the central halo mass densities.}
    \label{fig:all_density_4Mpc}
\end{figure*}

Fig.~\ref{fig:all_density_4Mpc} shows three different density estimates in one redshift bin with $\phot{\rho}$ in the top panels and $\sort{\rho}$ in the bottom panels. The left panels show number densities, the middle panels show stellar mass densities, and the right panels show halo mass densities only considering central galaxies. The colour and contours are proportional to the maximum bin value within each of the individual subplots. The dashed contour (in red) is set to a limit equal to the minimum contour level in the corresponding $\phot{\rho}$ subplot.

As expected, the photometric densities are underestimated in high-density regions. The high uncertainty of the photometric redshifts has the effect of smoothing out high- and low-density regions causing them to take on a more average density. After applying \SORT{}, low bias in the high-density regime is greatly improved and the distributions become more aligned with the line of equality. As shown by the dashed contour, \SORT{} struggles in the low-density regime, and the scatter is comparable to the photometric estimates. \SORT{} tends to overestimate its lowest local densities, which is a side effect of the clustering nature of the method. This is likely not something that can easily be remedied due to the simplicity of the \SORT{} method. By design, \SORT{} places galaxies near where it finds spectroscopic redshifts. Most spectroscopic redshifts will be in areas of higher density because this is where most galaxies are located. This tends to develop a cosmic structure that is highly clustered. \SORT{}'s ability to reconstruct low-density regions is dependent upon the quantity of high-quality redshifts found there, which will tend to be fairly limited.

\SORT{}'s estimates for central halo mass densities are not quite as good as its number and stellar mass densities, particularly for the mock CANDELS light cone where statistics are more limited (see Fig.~\ref{fig:all_density_4Mpc_small}). The likely cause of this is the removal of satellite galaxies. Halo masses for satellites are not tracked once they become sub-haloes and therefore were not considered in these calculations. The problem with this is that \SORT{} is effective on average for the \textit{full ensemble} of galaxies and does not discriminate different demographics (e.g., centrals versus satellites). We would not expect results to be as effective for a given subset of data since there is no mechanism within the method to treat different subsets differently. By removing satellites, we are decreasing the reliability of \SORT{}, particularly in high-density regions where most satellites reside. However, we note that despite this, \SORT{} still shows improvement over photometric density estimates.

\section{Discussion} \label{section:discussion}

\subsection{The Effects of Preserving the Rank Order} \label{section:control_test}
One of the key aspects of \SORT{} is the sorting itself. While it is clear that the reference sample provides a significant amount of information to \SORT{}, one might wonder what the effect of sorting is (in step~\ref{step:sorting} of Section~\ref{section:algorithm}). To test this, we ran a control algorithm that excluded step~\ref{step:sorting} where the rank ordering is done. The control results were computed simultaneously with the standard \SORT{} results and are identical in every way with the exclusion of the sorting; hence, the only difference in these two sets of results lies solely in the rank ordering. We call the results of this control algorithm $\ctrl{z}$.

The first result to consider is the $\Delta z$ histogram shown in Fig.~\ref{fig:dz_hist}. Performing a two-sided Kolmogorov--Smirnov test on $\Delta\sort{z}$ and $\Delta\ctrl{z}$ yields a $p$-value of $p<0.001$. This indicates with a high level of certainty that sorting indeed changes the distribution of $\Delta z$. To understand the differences, we consider the two features of $\Delta z$: the narrow, central peak and the broader base. 

\begin{figure*}
    \centering
    \includegraphics[width=\linewidth]{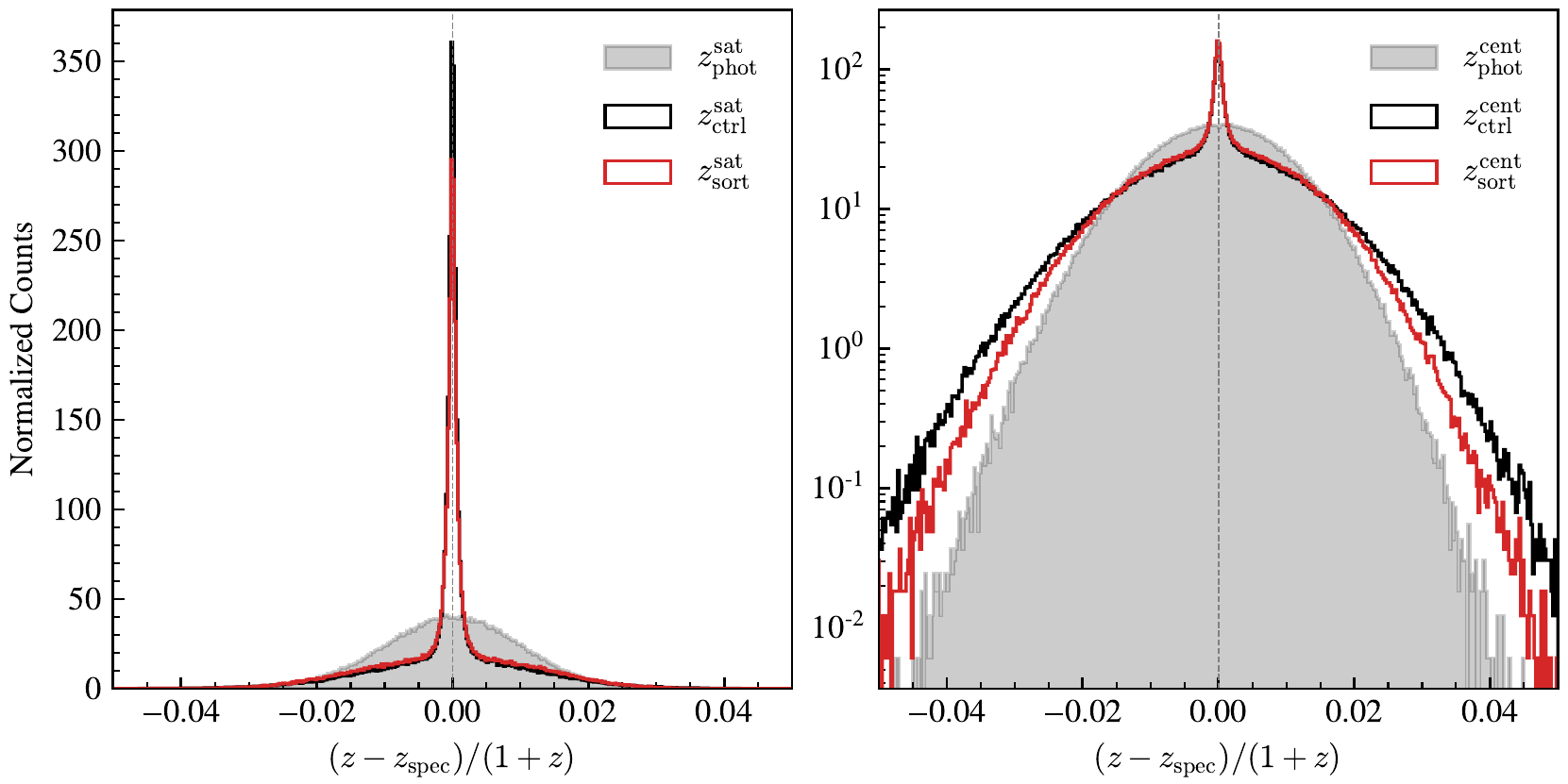}
    \caption{Normalized distribution of $\Delta z$ for $\sort{z}$, $\ctrl{z}$, and $\phot{z}$. The left panel shows the results using only satellite galaxies and the right panel shows the results using only central galaxies. See Section~\ref{section:control_test} for details. The left panel shows that the more highly-clustered results of $\ctrl{z}$ favor satellite galaxies. The right panel shows similar results between $\sort{z}$ and $\ctrl{z}$ at smaller errors; however, \SORT{} displays better treatment at larger errors ($\Delta z/(1+z)\gtrsim0.02$) with more rapidly declining tails.}
    \label{fig:dz_sort_vs_ctrl}
\end{figure*}

When looking at the peak around $\Delta z=\pm0.001$, we see the control sample outperforming \SORT{}. To interpret this result, we plotted the same diagram with the data broken into a set of central galaxies and a set of satellite galaxies, shown in Fig.~\ref{fig:dz_sort_vs_ctrl}. The left panel shows only satellites and the right panel shows only centrals. Looking at the peaks shows that the difference between $\Delta\sort{z}$ and $\Delta\ctrl{z}$ arises in the satellites. Satellite galaxies will tend to be situated more closely to their neighbors than a central galaxy. As such, the peak of $\Delta z^\text{sat}$ will tend to favor environments that are more densely packed. This is precisely what the control sample provides.

To illustrate this, consider some region of space containing a dense cluster of galaxies. If we assume there are $N$ galaxies along a pencil-beam-like sub-volume encompassing this dense cluster, we would expect each of those galaxies to have ${\sim}N$ recovered redshifts after \SORT{} is complete. In other words, since the cluster is dense, we expect most of the galaxies to fall within the sub-volumes of their neighbors. The sorting aspect of \SORT{} will always assign the lowest-redshift galaxies in this region the lowest recovered redshifts, and likewise assign the highest-redshift galaxies the highest recovered redshifts. This is simply following the condition laid out by stochastic ordering. Recalling that $\sort{z}$ is taken to be the median of a galaxy's assigned recovered redshifts, galaxies on the lower-redshift end of the cluster are biased to have a lower $\sort{z}$ and vice-versa at the higher-redshift end. In contrast to this, the control sample has no such bias. Each of the galaxies in the cluster will receive a random recovered redshift. After the algorithm completes, each galaxy will have a mixture of ${\sim}N$ high and low recovered redshifts which will tend to have a median towards the centre of the cluster. This centralization makes the cluster more dense than \SORT{} would make it, thus favoring $\Delta z^\text{sat}$.

Let us now consider the right panel of Fig.~\ref{fig:dz_sort_vs_ctrl}, $\Delta z^\text{cent}$. In this case, we have observed that there is no appreciable difference between $\Delta\sort{z}$ and $\Delta\ctrl{z}$ when it comes to the peak. However, if we look beyond the peak, we can see that \SORT{} performs better than the control sample. $\Delta\sort{z}$ tends to have higher counts than $\Delta\ctrl{z}$ up until the point where the tails of their distributions become broader than that of $\Delta\phot{z}$, around $\Delta z^\text{cent}=\pm0.015$. Beyond this point, $\Delta\sort{z}$ has a steeper distribution, signifying its better overall recovery of redshift estimates. This relative shape is also present for the $\Delta z$ histogram of the entire set of galaxies, though difficult to see in Fig.~\ref{fig:dz_hist}.

To further investigate the effects of sorting, we can consider the \tpcf{}. This metric provides a better characterization of the full three-dimensional distribution of galaxies than $\Delta z$. Fig.~\ref{fig:2pcf_3styles} shows a clear distinction between $\sort{z}$ and $\ctrl{z}$. \SORT{} is able to accurately reproduce $\spec{\xi}$(s) on scales of $s\gtrsim2.5~h^{-1}$Mpc. Due to the centralization and higher density produced by the control sample, $\ctrl{\xi}$(s) ends up on average around 25 per cent higher than $\sort{\xi}$(s). In other words, $\ctrl{z}$ is overestimating the clustering while $\sort{z}$ is not.

We conclude that while much of the information is provided by the reference sample, the sorting aspect of \SORT{} certainly provides useful information as well. This information is most evident when considering the \tpcf{} where the lack of sorting leads to over-clustering by around 25 per cent. The only drawback to sorting comes with the $\Delta z$ histogram of satellite galaxies. This is a difficult issue to resolve because \SORT{} uses one prescription to treat two distinct demographics, and information about which galaxies are centrals or satellites is not readily available for real observations.

\subsection{Sub-volume Parameters} \label{section:selection_volume}
The \SORT{} parameters $N^\text{min}_\text{ref}$, $R$, and $\Delta_z$ determine the sizes of the cylindrical sub-volumes that surround each galaxy during the \SORT{} procedure. A balance must be struck for these parameters in order for \SORT{} to produce reasonable results.

Having a larger cylinder radius allows for more of the environment to be taken into consideration when looking for reference galaxies. This can be useful in cases where galaxies are near the outer edge of a large cluster of galaxies. If the radius is too small, the inner region of the cluster may not be detected by the pencil-beam-like sub-volume. This leaves the galaxy more susceptible to being pulled toward denser regions that may be close on the sky but not in redshift. On the other hand, making the radius too large can also be problematic. With a large radius, galaxies that are not particularly nearby on the sky, but still within the sub-volume, will be pulled toward the redshifts of denser regions. Because \SORT{} only moves galaxies along the line of sight, these galaxies will be placed around the same redshift as another group of galaxies but with a seemingly `incorrect' position on the sky. The result is a distribution of galaxies that becomes elongated in a plane perpendicular to the line of sight.

This effect can be seen in Fig.~\ref{fig:square_field} and Fig.~\ref{fig:square_field_alt} and is a signature of the \SORT{} method. In low-density regions (e.g., the upper right corner of the $\sort{z}$ panel), we see horizontal formations of galaxies. The magnitude of this effect can be limited by adjusting $R_\text{max}$ or $N^\text{min}_\text{ref}$. As $N^\text{min}_\text{ref}$ becomes smaller, the radius of the average sub-volume will also be smaller, leading to narrower horizontal formations. This may come at the expense of \SORT{}'s overall performance, though. We found that increasing $N^\text{min}_\text{ref}$ from two to four provided a better estimate of the \tpcf{}, for example. If, instead, $R_\text{max}$ is adjusted, one must take care to not make it too small relative to $N^\text{min}_\text{ref}$. If $R_\text{max}$ is too small, the fail rate of the \SORT{} algorithm will increase as the maximum sub-volume size is too constrained to find enough reference galaxies. Likewise, the fail rate will also increase if $N^\text{min}_\text{ref}$ is too large for a given sub-volume size.

The redshift cut imposed by $\Delta_z$ is a new addition to \SORT{}. In the original \SORT{} paper \citep{2018MNRAS.473..366T} which looked at nearby galaxies in a wider, shallower field, an apparent magnitude cutoff was imposed such that only galaxies within $\pm\delta m$ of the $i$th galaxy were considered in the sub-volume. When \SORT{} is applied to a deeper field, the magnitude cut is not sufficient to allow \SORT{} to perform well. The same range of magnitudes can be found at opposite ends of the light cone, which leaves too large of a range of possible recovered redshifts for a given galaxy.

To limit the range of redshifts that are considered neighbors of the $i$th galaxy, a redshift cut based on a galaxy's photometric redshift was implemented to replace the magnitude cut. The length chosen for the cylinder should be based on the photometric uncertainty. In this case, we have assumed the photometric uncertainties are Gaussian. As such, we have chosen $\Delta_z=2.5\sigma_z^\text{ph}$ to allow the majority of photometric galaxies the potential of recovering their true redshift. The value of this parameter was not thoroughly tested, however, and may not be optimal. We emphasize that this parameter, as well as other \SORT{} parameters, should be tested to find optimal values when applied to different surveys. The values used in this paper correspond to sensible length scales, but optimal values will likely vary depending on the metric one wishes to optimize.

\subsection{Limitations}
One main limitation of the \SORT{} method is its dependence on a reference sample. The limitations of this dependence are twofold. First, there is a limitation to the length scale that \SORT{} will be able to properly recover. Dispersion velocities of $v\approx200$~km~s$^{-1}$ correspond to lengths of $\sim$4--6~$h^{-1}$Mpc for $z=1$--2. This is a rough limit of \SORT{}'s accuracy relative to the \textit{true} distribution of galaxies (i.e. not the spectroscopic distribution to which results were compared in this paper) due to redshift-space distortions. Second, \SORT{} requires a structural outline by the reference sample to recover an accurate distribution of galaxies. If the reference sample does not outline a feature of the cosmic web, then \SORT{} will not be able to reconstruct it. This effect is most significant in low-density regions. The fraction of galaxies found in low-density regions will naturally be low. An even lower fraction of those galaxies will be reference galaxies. Without reference galaxies, \SORT{} will not be able to reproduce an accurate distribution of galaxies in these regions.

A second limitation of the \SORT{} method is the fact that it can only improve redshift estimates collectively for ensembles of galaxies. Fig.~\ref{fig:dz_hist} shows a tall peak surrounding $\Delta z=0$ where a significant fraction of redshifts have been improved, but there is no way to tell which galaxies are in this peak or which galaxies are in the tails. Despite \SORT{} doing a fairly good job of recovering the large-scale structure of galaxies, there will undoubtedly be some galaxies placed in the wrong environments. Higher accuracy redshift estimates are still required to properly place galaxies on an individual basis. However, as shown in the previous section, \SORT{} can still be used to infer local densities (particularly, average or higher densities) reasonably well.

\subsection{Future Considerations} \label{section:future_considerations}
One possible next step for \SORT{} is updating the assignment of spectroscopic redshifts to create the reference sample. In this paper, spectroscopic redshifts were assigned randomly to 10 per cent of galaxies within three complete redshift bins. To make tests of the \SORT{} method more realistic, one could model the reference galaxy selection using the methods chosen by large imaging surveys (e.g., selecting a mixture of brighter galaxies and galaxies with high star formations rates that produce strong emission lines).

Another possible step is to improve redshift assignments within sub-volumes as the \SORT{} method is carried out. There are currently no considerations given to the angular correlations within each sub-volume. As shown in the left panel of Fig.~\ref{fig:dz_sort_vs_ctrl}, the current treatment of satellite galaxies by \SORT{} is not optimal.\footnote{That is not to say that $\ctrl{z}$ \textit{is} optimal, but it is enough to demonstrate that \SORT{} is not.} This could potentially be improved by assigning similar recovered redshifts to galaxies that appear highly clustered on the sky. This could also be implemented when determining the final $\sort{z}$ redshift of a galaxy. The final selection from a galaxy's pool of recovered redshifts at the end of the method could be biased to redshifts where the galaxy appears more clustered on the sky, as opposed to taking a simple median. Such considerations could also prove useful in reducing the horizontal structures produced by \SORT{} discussed in Section~\ref{section:selection_volume}.

Yet another improvement would be to treat satellite galaxies more realistically than we have done, as described in Appendix \ref{appendix:satellite_coords}. For example, an improved semi-analytic treatment of satellite galaxies could be based on the recent \texttt{SatGen} papers \citep{2021MNRAS.502..621J, 2021MNRAS.tmp.2848G, 2021MNRAS.508.2944G}.

Recently, it has been shown that correlations between galaxy and halo properties create observable signatures in local environments \citep{2021arXiv210105280B}. In particular, halo spin, concentration, growth rate, and interaction history have all been shown to leave scale-dependent signatures in both \tpcf{}s and the distributions of distances to galaxies' $k$th nearest neighbors out to $z\sim2.5$. These determinations were based on projected two-dimensional environments so as to make them observationally accessible with low-resolution spectroscopy ($\sigma_z/(1+z)\lesssim0.005$). We have shown that \SORT{} is able to recover the full three-dimensional \tpcf{} as estimated with high-resolution spectroscopy. We also provide in Appendix~\ref{appendix:additional_figures} results using \SORT{} to estimate three-dimensional distances to $k$th nearest neighbors (see Fig.~\ref{fig:r_nn}). We expect \SORT{}'s ability to reasonably-well recover local environments may allow for further observationally-accessible detections of environmental signatures that result from galaxy--halo property correlations.

In this paper, we have not taken advantage of the fact that galaxy properties could depend on environment and/or location within the cosmic web. Indeed, it is well known that, for example, star forming / blue galaxies are less clustered than more quiescent / red galaxies \citep[see e.g.,][]{Li+2006,Zehavi+2011,Coil+2017,2019ApJ...884...76B,2021AJ....161...49B} and that more spheroid-like morphologies are more frequently in denser environments \citep[e.g.,][]{Dressler1980,Pearson+2021}. Thus, a natural next step within the framework of \SORT{} would be to divide the reference sample by galaxy properties. By doing so, it is expected that \SORT{} would be able to determine even more accurate redshifts than when not considering galaxy properties. 

Finally, we expect to test the performance of \SORT{} using real data sets from highly complete spectroscopic galaxy surveys, e.g., GAMA \citep{GAMADR3} and DESI \citep{2016arXiv161100036D}, in order to account for systematics that are present in real surveys but not properly modeled by our mock experiment. For instance, we expect that the so-called `catastrophic redshift' failures in photometric redshift methods will have a minor effect in the performance of \SORT{} as these are typically only a small fraction of the total sample. Other systematic differences include having non-Gaussian PDFs for the photometric redshifts and having a set of photometric galaxies with variable $\sigma^\text{ph}_z$ in the sample. For example, COSMOS2020 \citep{2021arXiv211013923W} obtained photometric redshift precision of $\sim$4 per cent for the faintest galaxies and better than 1 per cent for the brightest galaxies.

\section{Summary and Conclusions} \label{section:summary}
In this paper, we have tested the performance of the \SORT{} method \citep{2018MNRAS.473..366T} in mock high-redshift surveys. \SORT{} is a simple, efficient, and robust method that can be used to improve redshift estimates. It relies upon a reference sample of high-quality spectroscopic redshifts for which a precise distribution $\mathrm{d}N/\mathrm{d}z$ is known within pencil-beam-like sub-volumes of the survey. Within each sub-volume we:
\begin{enumerate}
    \item sample new `recovered' redshifts from the $\mathrm{d}N/\mathrm{d}z$ distribution of high-quality redshifts
    \item match the recovered redshifts one-to-one with the low-quality (photometric) redshifts such that the rank order is preserved.
\end{enumerate}
The second step is motivated by the fact that random variables drawn from Gaussian PDFs with equal, arbitrarily-large standard deviations satisfy stochastic ordering. In other words, if two redshift estimates $z_i$ and $z_j$ satisfy $z_i<z_j$, then their \textit{true} redshift values most likely satisfy $z_i^\text{true} \le z_j^\text{true}$. Thus, preserving the rank order makes the assigned recovered redshifts more likely to be close to their underlying true value. This process is repeated for sub-volumes surrounding each galaxy in the survey. The result is every galaxy with a low-quality redshift is assigned multiple recovered redshifts from which a new redshift estimate can be determined.

We ran the \SORT{} method on a wide-field 2 square degree mock light cone and a mock CANDELS light cone extracted from the Small MultiDark--Planck and Bolshoi--Planck $N$-body simulations, respectively, to test its performance in a pencil-beam-like survey spanning the redshift range $0.75<z<2.25$. After applying \SORT{}, we observe similar improvement from both mock catalogues and make the following determinations:

\begin{itemize}
    \item We observed overall improvement in redshift estimates, allowing for better reconstruction of the three-dimensional distribution of galaxies than photometric redshifts alone provide. This can be seen broadly in Fig.~\ref{fig:big_field} or more close up in Fig.~\ref{fig:square_field}.
    \item We also observed that \SORT{} produces much better agreement in the one-dimensional $\mathrm{d}N/\mathrm{d}z$ distribution (by design), allowing it to better identify large-scale structure along the line of sight as shown in Fig.~\ref{fig:z_dist1d}.
    \item Redshift errors with respect to spectroscopic estimates were significantly reduced for a fraction of galaxies at all redshifts within the light cone, while overall scatter was only moderately increased. One- and two-dimensional histograms of this are shown in Fig.~\ref{fig:dz_hist} and Fig.~\ref{fig:z_dist2d}, respectively.
    \item \SORT{} accurately recovers the spectroscopic redshift-space \tpcf{} down to scales of ${\gtrsim}2.5~h^{-1}$Mpc while photometric redshifts (with errors of $\sigma_z^\text{ph}/(1+z)=0.01$ corresponding to scales of $\sim$20--30~$h^{-1}$Mpc) drastically underestimate galaxy clustering. This is shown clearly in the top and bottom panels of Fig.~\ref{fig:2pcf_3styles}.
    \item \SORT{} is able to recover three-dimensional local densities in regions of average or higher density at scales of ${\gtrsim}4~h^{-1}$Mpc. Three different density histograms are shown in Fig.~\ref{fig:all_density_4Mpc}, and additional histograms are shown in Fig.~\ref{fig:all_density_1000kms} at a larger length scale of $l=2\frac{1000~\text{km~s}^{-1}}{c}(1+z)$.
\end{itemize}

We expect that such improved determinations of local galaxy environments will help to distinguish the effects of environmental properties (e.g., local density) on galaxy evolution from other effects, such as galaxy stellar or halo mass \citep[e.g.,][]{2010ApJ...721..193P,2013MNRAS.428.3306W,2020ApJ...890....7C,2021arXiv210105280B}.

\section*{Acknowledgements}
We thank Peter Behroozi for creating the infrastructure that supplied the halo catalogues used in this work and Doug Hellinger for help working with the catalogues. This work was partially based on data products created as part of the CANDELS Multi-Cycle Treasury Program under NASA contract NAS5-26555. ARP acknowledges financial support from CONACyT through `Ciencia Basica' grant 285721 and from DGAPA-UNAM through PAPIIT grant IA104118. AY is supported by an appointment to the NASA Postdoctoral Program (NPP) at NASA Goddard Space Flight Center, administered by Oak Ridge Associated Universities under contract with NASA. RSS acknowledges support from the Simons Foundation. We thank contributors to the Python programming language\footnote{\url{https://www.python.org/}}, SciPy\footnote{\url{https://www.scipy.org/} \citep{2020SciPy-NMeth}}, NumPy\footnote{\url{https://numpy.org/} \citep{2020NumPy-Array}}, Matplotlib\footnote{\url{https://matplotlib.org/} \citep{Hunter:2007}}, Astropy\footnote{\url{https://www.astropy.org/} \citep{astropy:2013, astropy:2018}}, and the free and open-source community.

\section*{Data Availability}
The mock CANDELS backward light cone used in this paper, as well as additional light cones for other CANDELS fields, are available at \url{https://www.simonsfoundation.org/candels-survey}.


\bibliographystyle{mnras}
\bibliography{main}



\appendix

\section{SORT Performance in a Mock CANDELS Light Cone} \label{appendix:candels_lc}
In addition to the 2 square degree wide-field light cone, \SORT{} was tested on a narrower $17\times41$ square arcmin light cone with 58,093 galaxies ($\sim$83 galaxies per square arcmin). All model parameters were kept the same, and we observed that \SORT{} continues to perform well at improving redshift estimates and determining local galaxy environments. Here we provide parallel results of the main text for \SORT{} applied to this light cone with minor adjustments detailed hereafter.

In Fig.~\ref{fig:z_dist2d_small}, we show the two-dimensional redshift histograms comparing $\sort{z}$ and $\phot{z}$ to $\spec{z}$. However, because statistics are more limited in the mock CANDELS light cone, all three redshift ranges of width $\Delta z=0.5$ were stacked on top of each other to create composite histograms. This allows recovery of similar distributions to those shown in Fig.~\ref{fig:z_dist2d}. In particular, the photometric redshifts maintain their Gaussian error distributions while \SORT{} redshifts build up along the line of equality.

We note that because \SORT{} is largely dependent upon $\reference{z}$, which comprises a small fraction of the total redshifts, \SORT{}'s ability to reconstruct the three-dimensional distribution of galaxies is susceptible to sample variance in $\reference{z}$. As such, for the mock CANDELS light cone where statistics are more limited, \SORT{} was run on the same light cone with ten different random seeds (which determine the selection of $\reference{z}$) to find an average result for the \tpcf{}. For each of the 10 seeds, an average \tpcf{} was calculated using 10 bootstraps. The averages of all 100 \tpcf{} estimates are shown in Fig.~\ref{fig:2pcf_3styles_small} for each of the redshift types in the range $0.75<z<1.25$. The error bars represent the standard deviations of $\xi(s)$ for each of the 10 seeds within each bin. We note that averaging over the 10 random seeds was done for all redshift types, though results will not vary much when using $\cosmo{z}$, $\spec{z}$, and $\phot{z}$. Due to the narrower geometry of this light cone compared to the wide-field light cone, the \tpcf{}s were only calculated out to 8~$h^{-1}$Mpc.

\begin{figure}
    \centering
    \includegraphics[width=\linewidth]{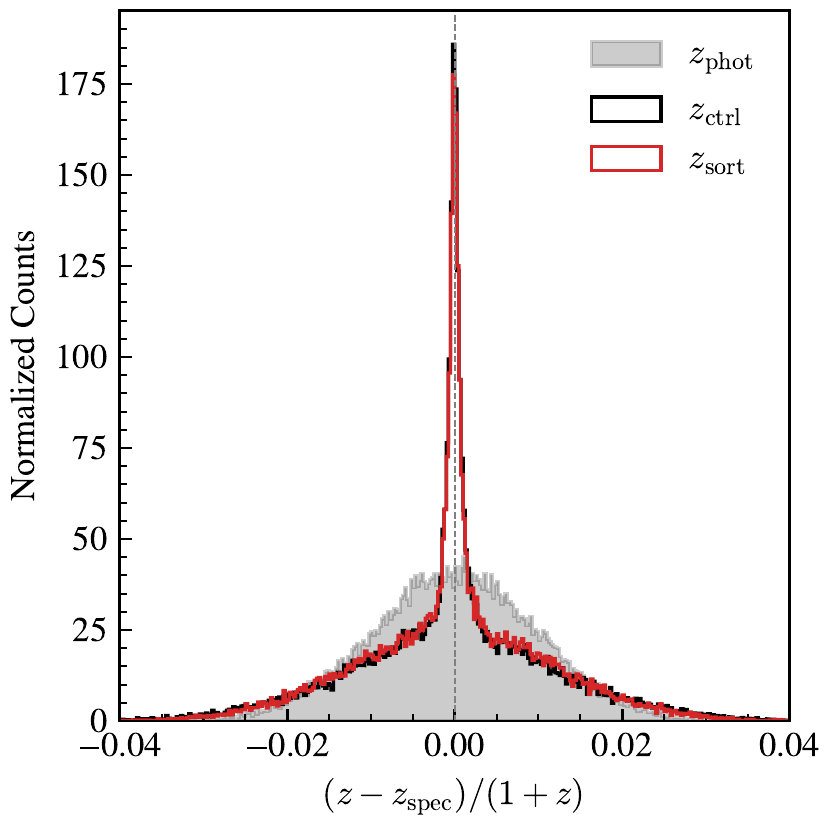}
    \caption{Normalized distribution of $\Delta z$ (excluding the spectroscopic sample) for $\sort{z}$, $\phot{z}$, and $\ctrl{z}$ using the mock CANDELS light cone (see Section~\ref{section:control_test} for details on $\ctrl{z}$). We recover a distribution of redshift errors similar to Fig.~\ref{fig:dz_hist} using the wide-field light cone. In particular, the $\Delta\sort{z}$ distribution is dominated by a tall, central peak of improved redshifts.
    }
    \label{fig:dz_hist_small}
\end{figure}

\begin{figure*}
    \centering
    \includegraphics[width=\linewidth]{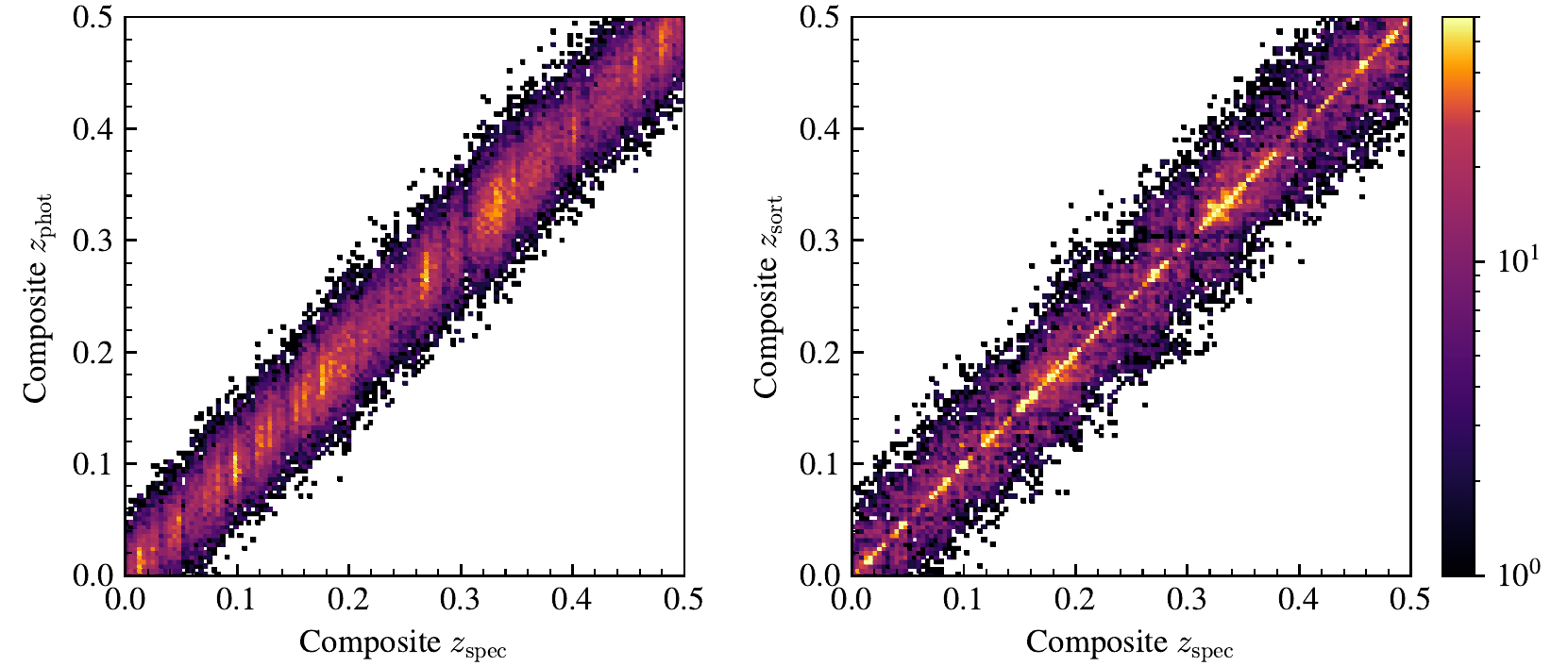}
    \caption{Two-dimensional redshift histograms for $\phot{z}$ and $\sort{z}$ relative to $\spec{z}$ with binning of 0.004 using the mock CANDELS light cone. The color bar shows the total number of counts in each bin. The data represent the full catalogue of redshifts broken into the three complete redshift bins of size $\Delta z=0.5$ that have been stacked on top of each other. In doing so, we are able to observe similar improvement in redshift estimates to the wide-field light cone after applying \SORT{}.}
    \label{fig:z_dist2d_small}
\end{figure*}

\begin{figure}
    \centering
    \includegraphics[width=\linewidth]{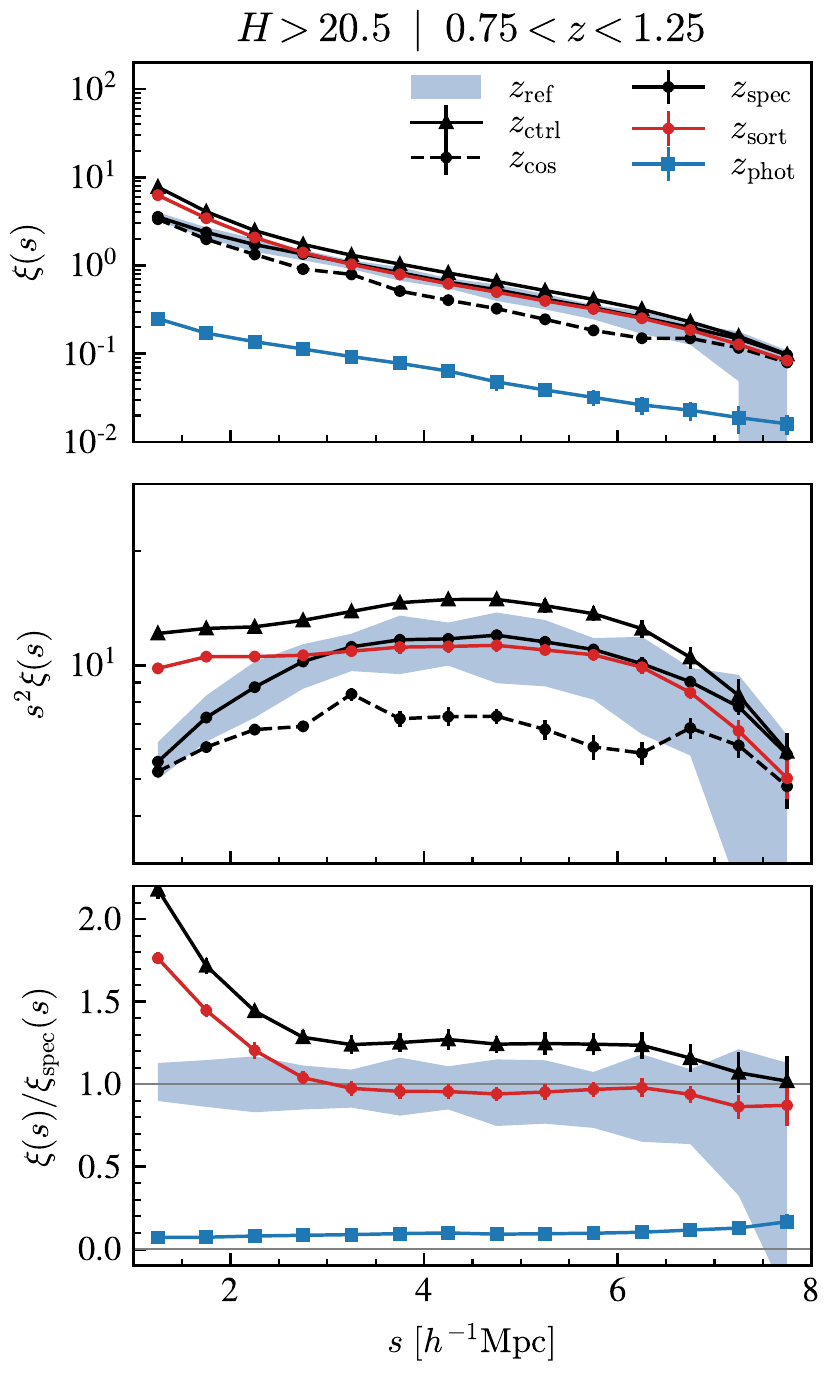}
    \caption{Two-point correlation functions (\tpcf{}s) of various redshift types as a function of redshift-space distance $s$ shown in three different ways using the mock CANDELS light cone. In each panel, the values plotted represent the mean result of running \SORT{} with 10 different random seeds, each bootstrapped 10 times, to average out sample variance when selecting the reference galaxies. We continue to see that $\phot{\xi}(s)$ is a poor estimate of the \tpcf{} and $\ctrl{z}$ overestimates the \tpcf{} while $\sort{\xi}(s)$ is accurate (relative to $\spec{\xi}(s)$) for $s\gtrsim2.5~h^{-1}$Mpc. See Section~\ref{section:control_test} for details on $\ctrl{z}$ (shown as black triangles.}
    \label{fig:2pcf_3styles_small}
\end{figure}

\begin{figure*}
    \centering
    \includegraphics[width=6in]{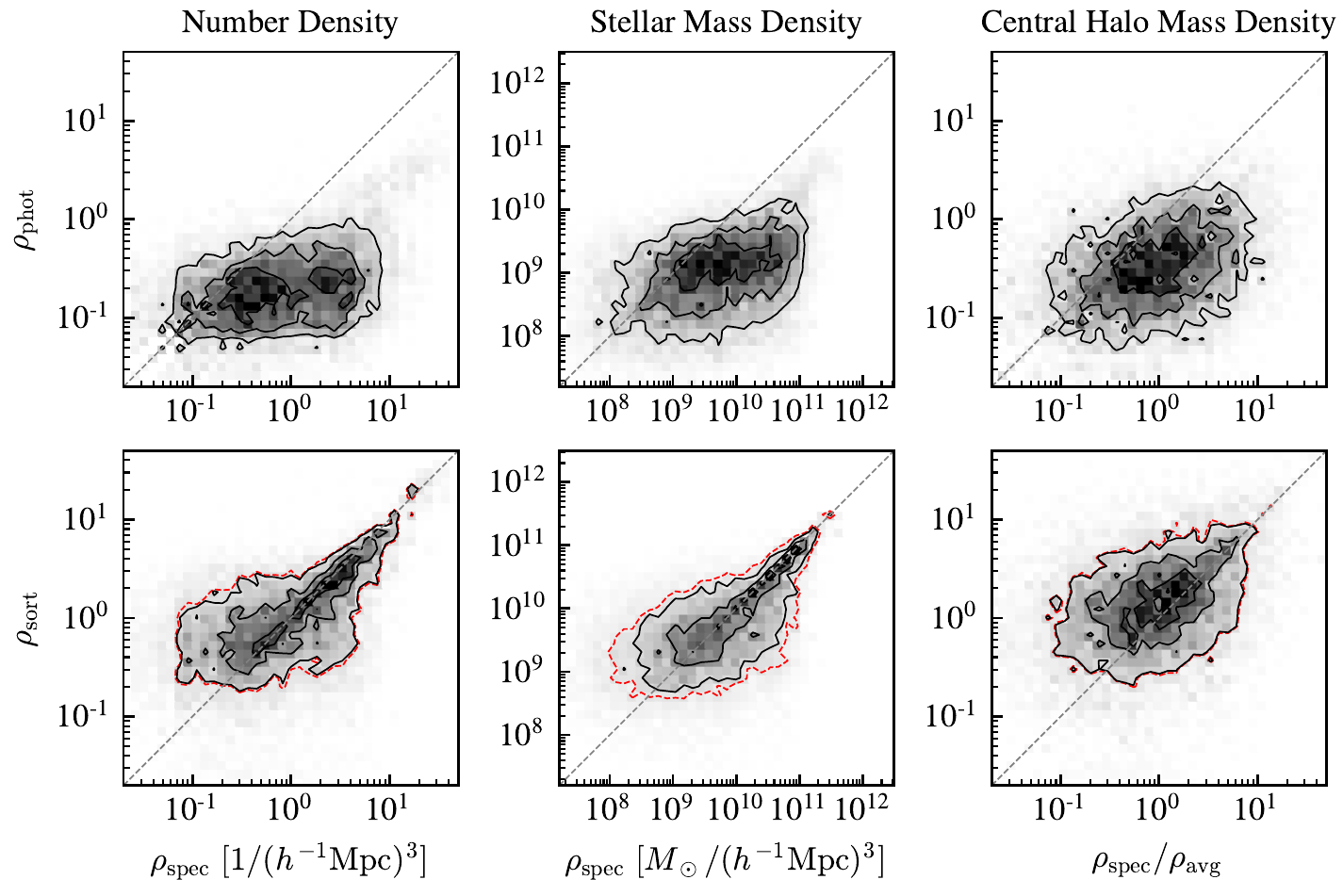}
    \caption{Two-dimensional density histograms for $\phot{\rho}$ and $\sort{\rho}$ in the range $0.75<z<1.25$ using the mock CANDELS light cone. The left panels show number densities, the middle panels show stellar mass densities, and the right panels show halo mass densities using only central galaxies. The solid contours represent limits of 25, 50, and 75 per cent of the maximum bin value in each subplot. The dashed contour (red) is set at a limit equal to the minimum contour level in the corresponding $\phot{\rho}$ subplot. Densities were calculated with a fixed cylinder length of 4~$h^{-1}$Mpc and a radius check this starting at ${\sim}0.5~h^{-1}$Mpc and expanding up to ${\sim}2~h^{-1}$Mpc as needed to encompass at least five galaxies. As with the wide-field light cone, we continue to see improvement in density estimates by \SORT{} compared to photometric estimates. See Section~\ref{section:density} for details on the central halo mass densities.}
    \label{fig:all_density_4Mpc_small}
\end{figure*}

\section{Assigning Three-Dimensional Coordinates to Satellite Galaxies} \label{appendix:satellite_coords}
The mock galaxy surveys used for this paper do not provide full three-dimensional coordinates for satellite galaxies. Instead, all satellite galaxies are assigned the same redshift as the dark matter halo they occupy. We used the following procedure to assign new three-dimensional coordinates and velocities to each of the satellites.

\subsection{Calculating the Position}
New positions were calculated for each satellite assuming that the satellites have the same radial  distribution as the dark matter \citep[see e.g.,][]{Berlind_Weinberg_2002,Cooray_Sheth_2002}.\footnote{Improved treatments based on observations are discussed in \cite{Watson+2012, 2018ARA&A..56..435W, Lange+2019, vdBosch+2019}.} The radial density profile was approximated using the NFW formula \citep{1996ApJ...462..563N, 1997ApJ...490..493N}
\begin{equation}
    \rho_\text{NFW}(r) = \frac{4\rho_s}{(r/R_s)(1+r/R_s)^2}.
\end{equation}
It is determined by two parameters, in this case $\rho_s$ and $R_s$. Alternatively, it can be determined by the halo mass, $\vir{M}$, and the halo concentration, $c_\text{vir}$, which is defined as
\begin{equation}
    c_\text{vir} = \frac{\vir{R}}{R_s}.
\end{equation}
The scale radius, $R_s$, is the radius at which the log-space derivative of $\rho_\text{NFW}(r)$ is -2. This could be found by fitting the NFW profile to each halo in the simulation. However, a more robust method is to find the Klypin scale radius using the $\vir{M}$--$V_\text{max}$ relation under the assumption of an NFW profile \citep{2011ApJ...740..102K}. The parameter $V_\text{max}$ is the maximum circular velocity of the halo (i.e. the maximum of $\sqrt{GM(r)/r}$, where $M(r)$ is the mass enclosed within a radial distance $r$). For an NFW profile, the maximum circular velocity occurs at $R_\text{max}=2.1626R_s$ \citep{2011ApJ...740..102K, 2013ApJ...762..109B}. With this, we calculated the Klypin concentration, $c_\text{vir,K}$, by numerically solving
\begin{equation}
        \frac{c_\text{vir,K}}{f(c_\text{vir,K})} = V_\text{max}^2\frac{\vir{R}}{G\vir{M}}\frac{2.1626}{f(2.1626)}
\end{equation}
where
\begin{equation}
    f(x) = \ln(1+x) - x/(1+x).
\end{equation}

The radial distribution for an NFW profile can also be written in terms of the halo's mass as
\begin{equation}
    M_h(r) = \vir{M}\times \vir{u}(r)
\end{equation}
where $\vir{u}(r)$ is
\begin{equation}
    \vir{u}(r) = \frac{\ln(1+c_\text{vir,K}x) - c_\text{vir,K}x / (1+c_\text{vir,K}x)} {\ln(1+c_\text{vir,K}) - c_\text{vir,K} / (1+c_\text{vir,K})}
\end{equation}
with $x=r/\vir{R}$. We can use this to sample new radial positions for the satellites within a halo. For each satellite in a given halo, the following procedure was followed.
\begin{enumerate}
    \item Generate three random numbers $U_r$, $U_\theta$, and $U_\phi$, each uniformly distributed between 0 and 1.

    \item Sample a radius from the distribution $\vir{u}(r)$. This can be done by finding the value $r$ such that $U_r-\vir{u}(r)=0$.

    \item Assign the new spherical coordinates $(r, \theta, \phi)$ to the satellite relative to the halo's centre where $\theta=\pi U_\theta$ and $\phi=2\pi U_\phi$.

    \item Assign new Cartesian coordinates $\mathbf{r}=(x,y,z)$ relative to the halo using
        \begin{align}
            &x = r\sin\theta\cos\phi\\
            &y = r\sin\theta\sin\phi\\
            &z = r\cos\theta.
        \end{align}

    \item Get the position of the satellite relative to the box of the simulation using $\mathbf{R} = \mathbf{R}_\text{h} + \mathbf{r}$, where $\mathbf{R}_\text{h}$ is the position of the halo relative to the simulation box. A cosmological redshift, $\cosmo{z}$, can be inferred from the new satellite position.
\end{enumerate}

\subsection{Calculating the Velocity}
To find the line-of-sight redshift, $z_\text{los}$, of each satellite, we must account for the effects of the peculiar velocity along the line of sight. The peculiar velocity of each satellite will depend on its position within a halo. Using the new radial distribution of satellites, as well as properties of the haloes within which the satellites reside, we can estimate new satellite velocities.
\begin{enumerate}
    \item By assuming that the satellite velocities trace the dark matter particle velocities within an NFW halo, we can calculate the velocity dispersion of the satellites at a distance $r$ from the halo's centre using
        \begin{equation}
            \sigma^2(r) = \frac{c_\text{vir,K}\vir{V}^2}{\mu(c_\text{vir,K})}{\frac{r}{R_s}}\parenth{1+\frac{r}{R_s}}^2 \int_{r/R_s}^\infty\frac{\mu(x)\mathrm{d}x}{x^3(1+x)^2}
        \end{equation}
    where $\mu(x)$ is defined as
        \begin{equation}
            \mu(x) = \ln(1+c_\text{vir,K}x) - c_\text{vir,K}x/(1+c_\text{vir,K}x).
        \end{equation}
    
    \item Sample a velocity $v$ from the Gaussian distribution
        \begin{equation}
            P(v) = \frac{1}{\sqrt{2\pi\sigma^2(r)}}\exp\parenth{-\frac{v^2}{2\sigma^2(r)}}.
        \end{equation}
    \item Generate two random numbers $U_\theta$ and $U_\phi$, each uniformly distributed between 0 and 1.

    \item Using $\theta=\pi U_\theta$ and $\phi=2\pi U_\phi$, the components of the satellite's velocity vector $\mathbf{v} = (v_x, v_y, v_z)$ relative to the halo's centre are
        \begin{align}
            &v_x = v\sin\theta\cos\phi\\
            &v_y = v\sin\theta\sin\phi\\
            &v_z = v\cos\theta.
        \end{align}
    \item With respect to the box of the simulation, the satellite's velocity is $\mathbf{V} = \mathbf{V}_\text{h} + \mathbf{v}$, where $\mathbf{V}_h$ is the halo's velocity with respect to the simulation box.
\end{enumerate}
The component of the velocity along the line of sight can be found by the new position and velocity vectors:
\begin{equation}
    v_\text{los} = \mathbf{V} \boldsymbol{\cdot} \mathbf{\hat{R}}
\end{equation}
where $\mathbf{\hat{R}}$ is the unit vector pointing to the satellite's position. The final redshift can be calculated using
\begin{equation}
    z_\text{los} = \cosmo{z} + \frac{v_\text{los}}{c}(1+\cosmo{z}),
\end{equation}
where $c$ is the speed of light.

\section{SORT Performance with Larger Photometric Uncertainties} \label{appendix:sigma_ph_0.02}
Our fiducial photometric uncertainty is somewhat optimistic at $\sigma_z^\text{ph}/(1+z)=0.01$, though not entirely unrealistic as future redshift estimates are expected to have photometric uncertainties of $\sigma_z^\text{ph}/(1+z)\approx0.02$ or better. Nevertheless, here we present brief results of \SORT{} for larger photometric uncertainties.

\begin{figure}
    \centering
    \includegraphics[width=\linewidth]{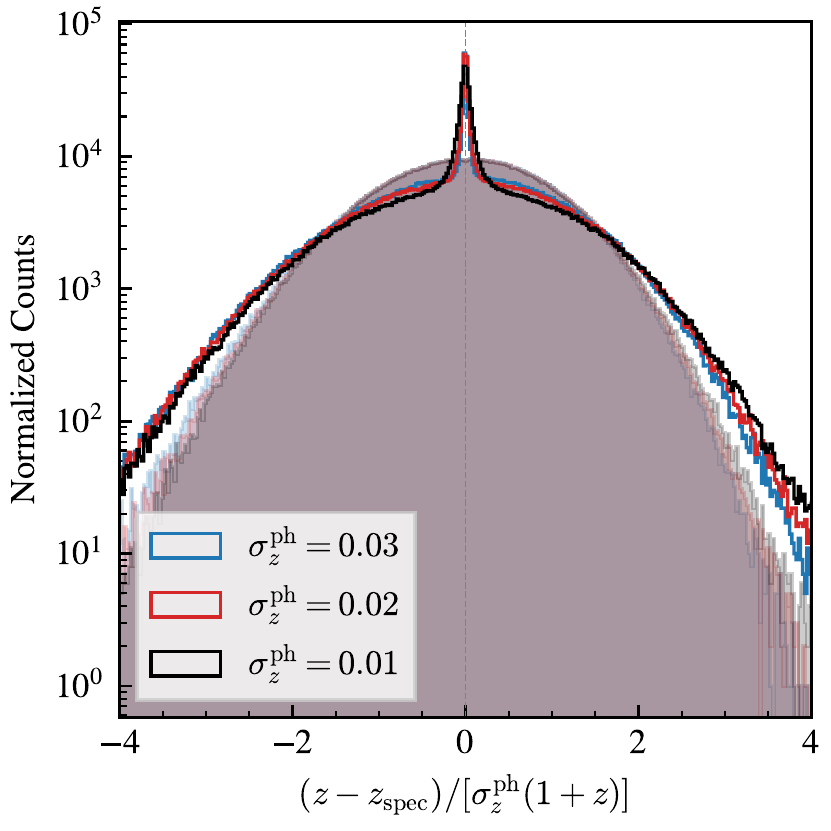}
    \caption{Normalized distributions of $\Delta z$ (excluding the spectroscopic sample) for $\sort{z}$ and $\phot{z}$ using three different photometric uncertainties. The distributions are normalized by their respective photometric uncertainties. We observe that the \textit{relative} improvement of redshifts by \SORT{} with respect to a given $\sigma_z^\text{ph}$ is generally independent of $\sigma_z^\text{ph}$. There are some dissimilarities, however. In particular, asymmetry in the tails of the distributions grows with $\sigma_z^\text{ph}$, and the width of the peak decreases with increasing $\sigma_z^\text{ph}$.}
    \label{fig:dz_pherrs_avg}
\end{figure}

We reiterate that stochastic ordering holds true for Gaussian PDFs with arbitrarily-large standard deviations. We should therefore expect to see similar redshift improvement when increasing the photometric uncertainty. The results for $\Delta z$ are shown in Fig.~\ref{fig:dz_pherrs_avg} with the fiducial results in black and the results with higher photometric uncertainties in red and blue. The histograms have been normalized by the photometric uncertainty to show the \textit{relative} performance of \SORT{} as $\sigma_z^\text{ph}$ is increased. To deal with biases from sample variance in $\reference{z}$, the histograms show the collection of \SORT{} results using ten different random seeds. We observe that \SORT{}'s improvement of redshifts with respect to a given photometric uncertainty remains largely unchanged as $\sigma_z^\text{ph}$ increases. In all three cases we see the same general features: (i) a similar overall standard deviation in $\Delta\sort{z}$ and $\Delta\phot{z}$, (ii) a modest increase in scatter at the tail ends of $\Delta\sort{z}$ compared to $\Delta\phot{z}$, and (iii) a tall central peak of improved redshifts.

The two most notable differences are an increase in asymmetry in the tails of the histograms and a decrease in peak width as $\sigma_z^\text{ph}$ increases. The asymmetry of the tails will be mostly irrelevant to the net result of \SORT{} as the counts are around two orders of magnitude lower than the peak which dominates the distribution. Though we do not test this here, the width of the peak is likely more relevant to the final results of \SORT{}. However, even with a photometric uncertainty of $\sigma_z^\text{ph}/(1+z)=0.02$, \SORT{} is still able to fairly well recover the \tpcf{} at similar scales of $s\gtrsim2.5~h^{-1}$Mpc (shown in the bottom panel of Fig.~\ref{fig:2pcf_3styles}). 

\section{Note on Making Figures With Many Points for Publication} \label{appendix:figure_code}
This paper features several figures that contain of order thousands of points (e.g., Fig~\ref{fig:big_field}). One challenge when dealing with figures such as these is keeping them clear and vectorized while simultaneously managing the size of the file. One option is to rasterize the entire figure by saving it as a jpg or png. However, this sacrifices the vectorization of the axes and labels, which are not the cause of file size issues. The result is an overall blurry figure when inspected closely. We provide in Fig.~\ref{fig:sample_code} some example code (using Matplotlib in Python) that demonstrates a better solution the reader may find useful.

\begin{figure}
    \centering
    \begin{minted}{python}
    import matplotlib.pyplot as plt
    import numpy as np
    
    x = np.linspace(0, 10, 50000)
    y = x + np.random.normal(0, 1, len(x))
    
    fig, ax = plt.subplots(dpi=100)
    ax.scatter(x, y, s=1, rasterized=True)
    ax.plot([0, 10], [0, 10], color='k')
    fig.savefig('figure.pdf')
    \end{minted}
    \caption{Sample code for producing figures with individually rasterized elements.}
    \label{fig:sample_code}
\end{figure}

Each plotted element on the figure can individually be rasterized while still maintaining vectorized axes and labels (or other plotted elements, e.g., lines, histograms, etc). The dpi for the rasterized elements can be controlled when the figure is created. In this particular example, rasterizing the scattered points lowers the file size by over an order of magnitude.

\section{Additional Figures} \label{appendix:additional_figures}
\begin{figure}
    \centering
    \includegraphics[width=\linewidth]{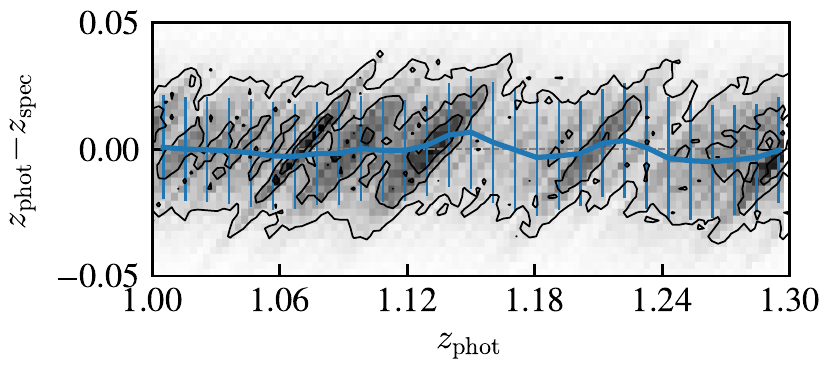}
    \includegraphics[width=\linewidth]{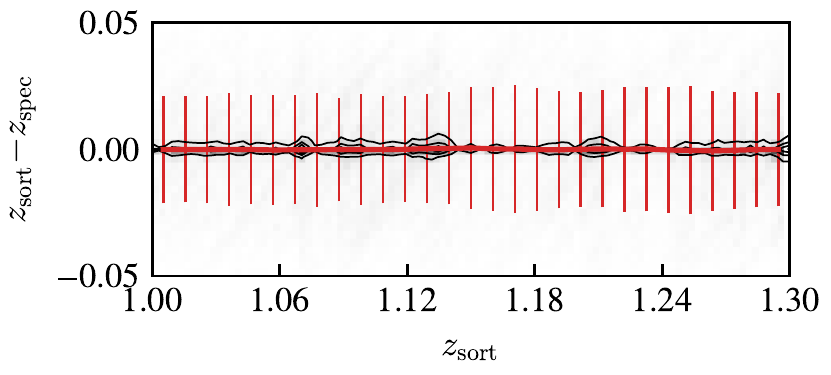}
    \caption{Normalized two-dimensional histograms for errors in $\phot{z}$ and $\sort{z}$ relative to $\spec{z}$. The contours show the limits where counts are at least 25, 50, or 75 per cent of the maximum value in each of the two subplots. While only the range $z=1$--1.3 is shown, the results are representative of the entire light cone. There is a clear bias in the error of $\phot{z}$ in regions of higher density. This bias is shown as a blue line which designates the median value of all redshifts within a series of bins along with $1\sigma$ error bars. As with Fig.~\ref{fig:dz_hist}, both $\Delta\phot{z}$ and $\Delta\sort{z}$ have similar standard deviations. After applying \SORT{}, though, the error bias is almost completely removed for the entire redshift range of the light cone.}
    \label{fig:z_diff}
\end{figure}

\begin{figure}
    \centering
    \includegraphics[width=\linewidth]{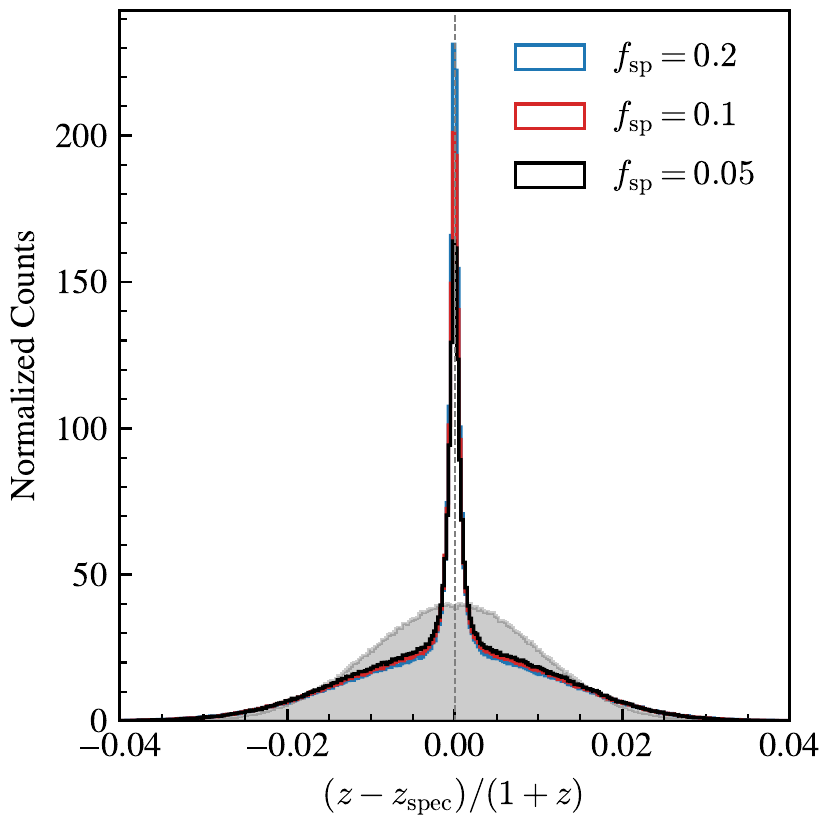}
    \caption{Normalized distribution of $\Delta z$ (excluding the spectroscopic sample) for $\phot{z}$ and $\sort{z}$ using three different spectroscopic fractions. As the spectroscopic fraction increases, \SORT{} produces a taller peak surrounding $\Delta z=0$. Even with a spectroscopic fraction as low as 5 per cent, \SORT{} still improves redshift estimates for a significant fraction of galaxies. The efficiency of \SORT{} is rooted in the fact that most galaxies will tend to occupy a relatively small volume. Therefore it only takes a relatively small fraction of galaxies to reasonably trace the underlying distribution.}
    \label{fig:dz_hist_fg}
\end{figure}

\begin{figure}
    \centering
    \includegraphics[width=\linewidth]{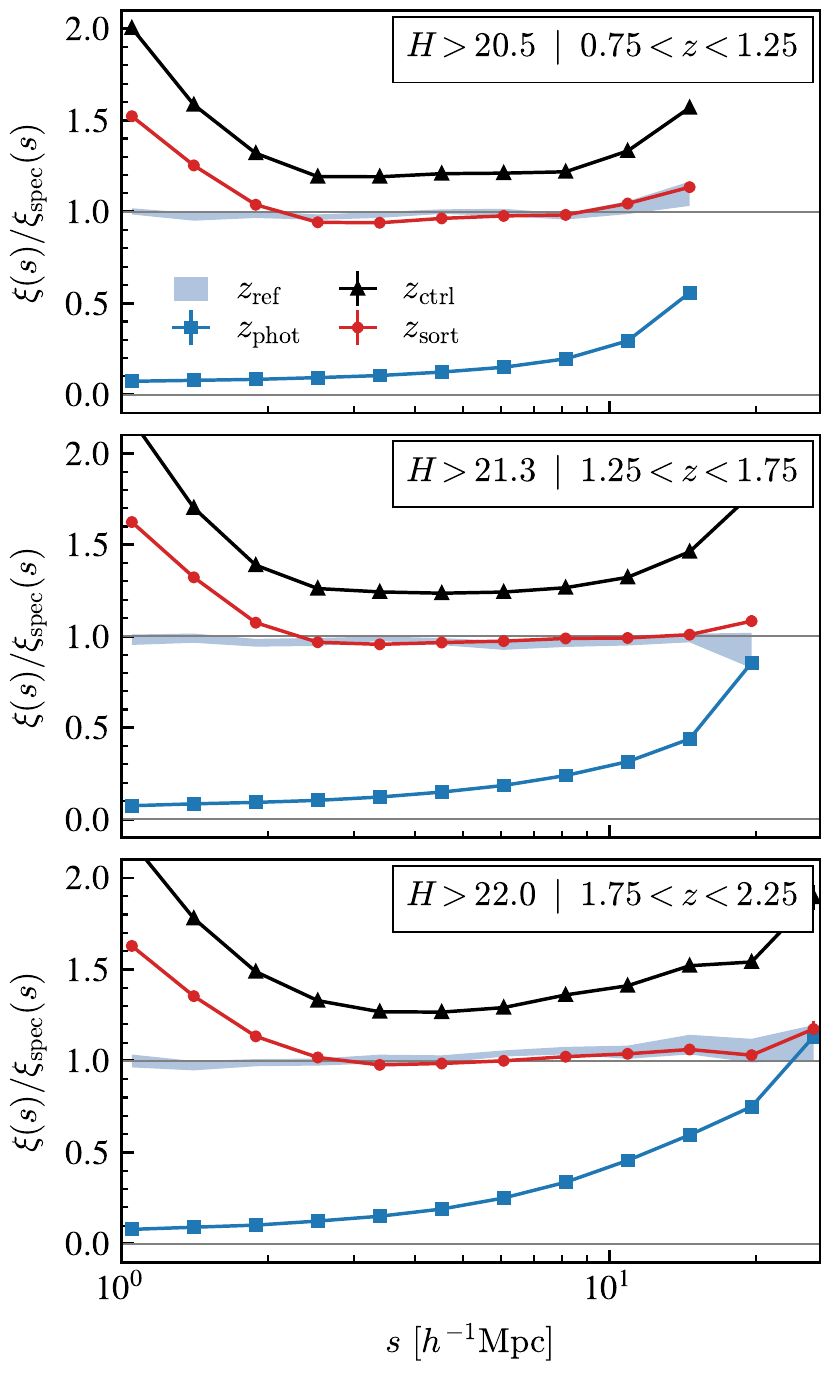}
    \caption{Two-point correlation function (\tpcf{}) ratios using $\reference{z}$, $\phot{z}$, $\sort{z}$, and $\ctrl{z}$ with respect to $\spec{z}$ as a function of redshift-space distance $s$ in three complete redshift bins. The \tpcf{}s were calculated out to distances of ${\sim}18$--$30~h^{-1}$Mpc, limited by the sizes of each redshift bin. The results show the mean value of the \tpcf{}s along with 1$\sigma$ error bars after 10 bootstrap samples. Note that the error bars are too small to be seen. The \tpcf{} estimates provided by $\sort{\xi}(s)$ show significant improvement over $\phot{\xi}(s)$ and accurately recover $\spec{\xi}(s)$ at scales of $s\gtrsim2.5~h^{-1}$Mpc. We also observe the continued trend of $\ctrl{\xi}(s)$ overestimating the \tpcf{} at all scales relative to $\sort{\xi}(s)$ (see Section~\ref{section:control_test} for details).}
    \label{fig:2pcf_full_range}
\end{figure}

\begin{figure*}
    \centering
    \includegraphics[width=\linewidth]{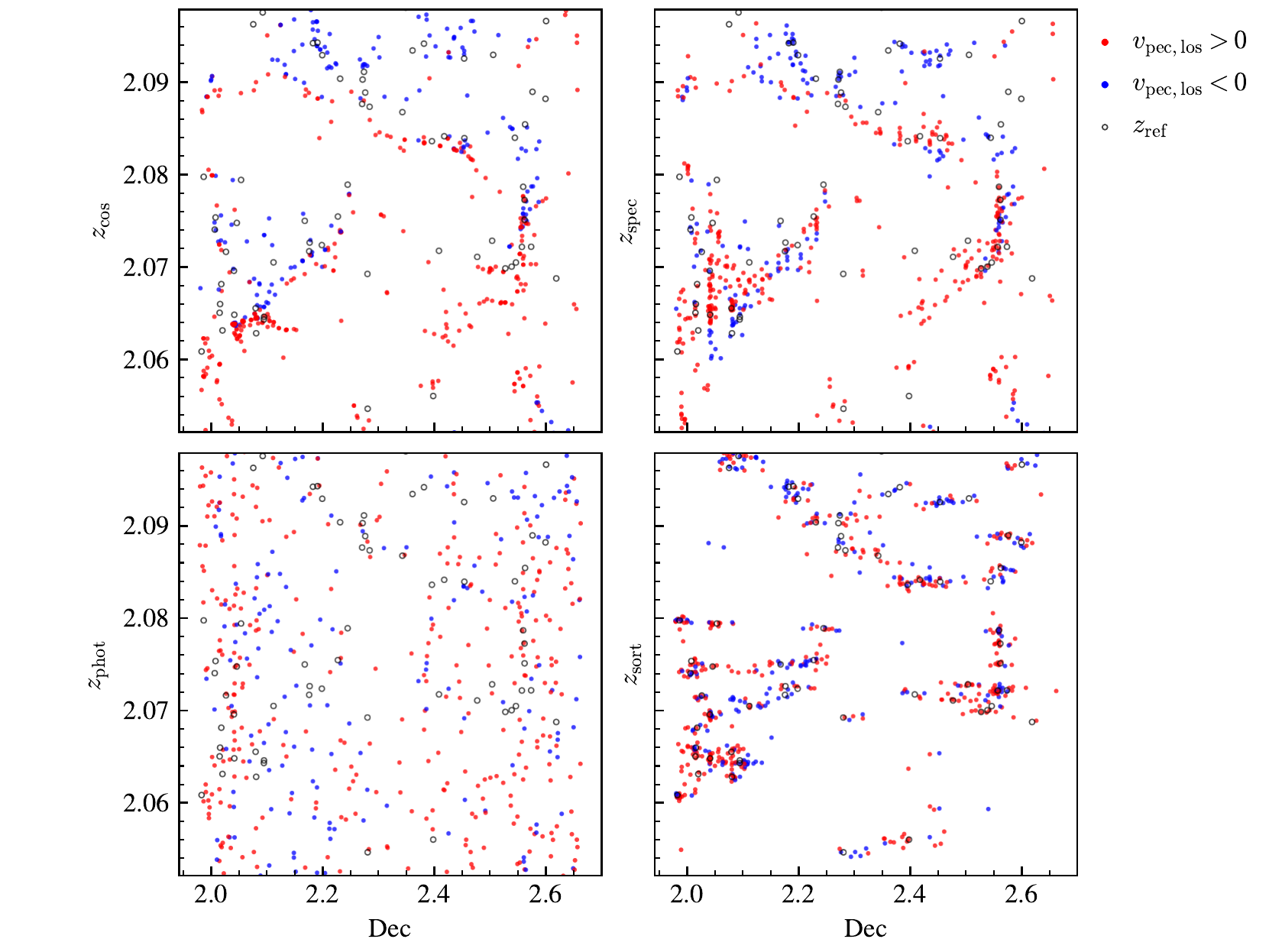}
    \caption{Right ascension slices (thickness 0.1$^\circ$) of galaxy distributions using different redshifts in a roughly $44\times44~h^{-1}$Mpc region of space. The red and blue colouring denotes the direction of the peculiar velocity along the line of sight (red is positive and blue is negative). The black rings with empty centres are reference galaxies. Using the outline of the reference galaxies, \SORT{} is able to recover the distinctive features in this region -- in particular, the large filamentary structure across the top and right side of the panels, as well as the more dense group of galaxies in the lower left. We also note the presence of a characteristic feature found in \SORT{} galaxy distributions -- namely, horizontal rows of galaxies where there are few reference galaxies. In these areas, the radii of the sub-volumes within which \SORT{} searches must expand to find reference galaxies. Galaxies are then pulled along the line of sight to an incorrect redshift, creating elongated features in a plane perpendicular to the line of sight. See Section~\ref{section:selection_volume} for details.}
    \label{fig:square_field_alt}
\end{figure*}

\begin{figure*}
    \centering
    \includegraphics[]{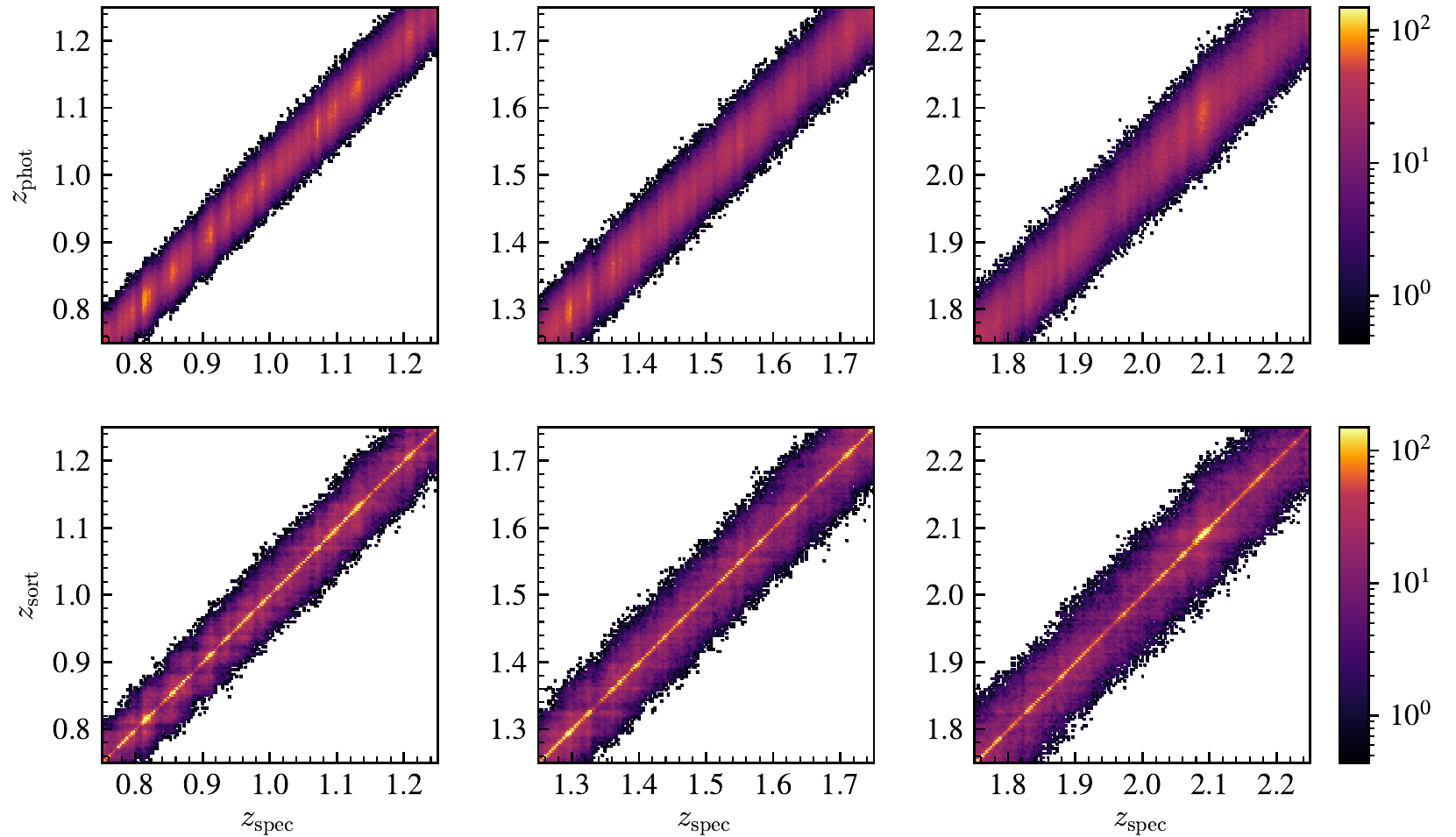}
    \caption{Normalized two-dimensional redshift histograms for $\phot{z}$ and $\sort{z}$ compared to $\spec{z}$ in all redshift bins. The $\sort{z}$ distributions show significant improvement as counts build up along the line of equality while overall scatter for larger redshift errors is only modestly increased. This effect is consistent across all redshift ranges.}
    \label{fig:individual_z_dist2d}
\end{figure*}

\begin{figure*}
    \centering
    \includegraphics[width=6in]{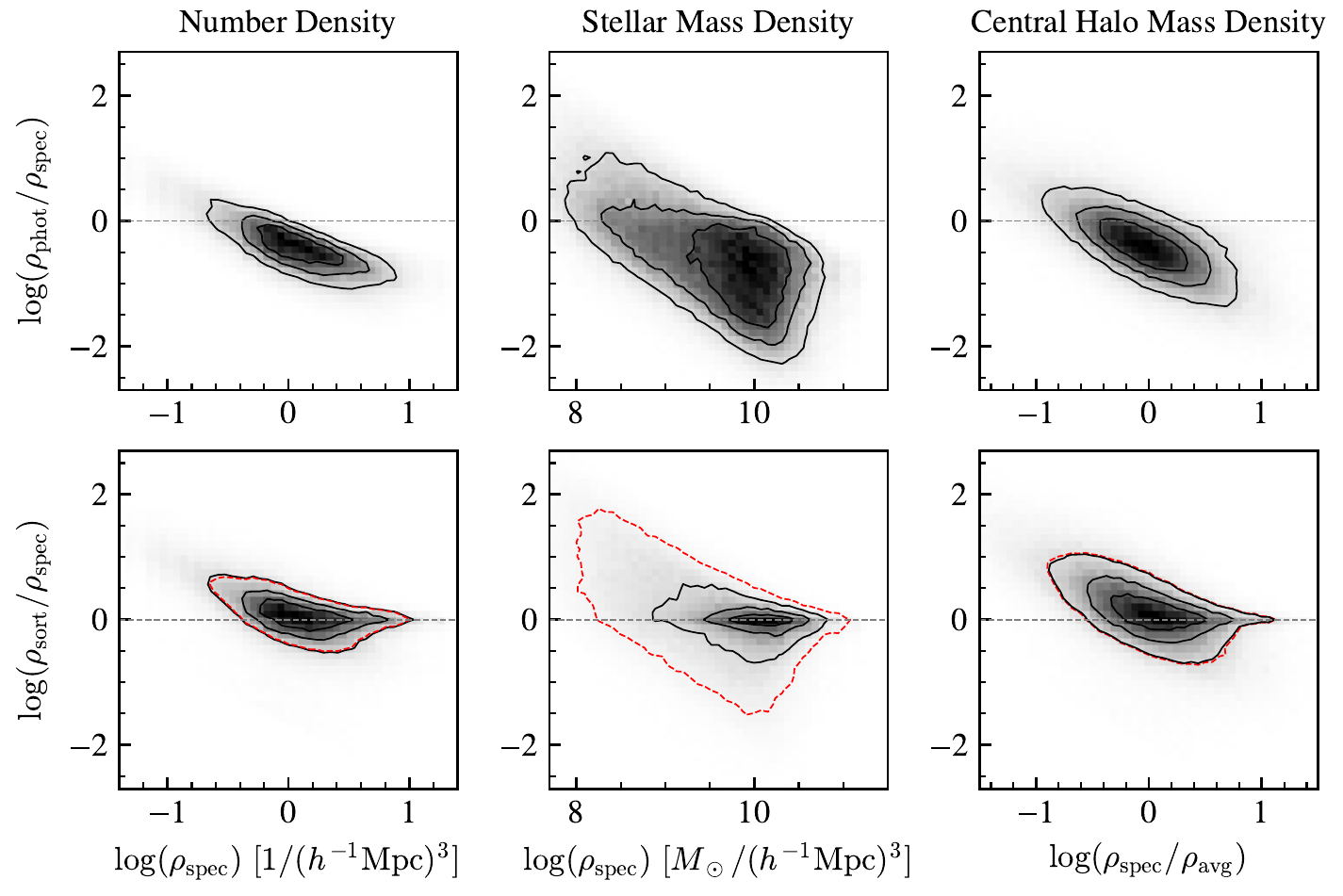}
    \includegraphics[width=6in]{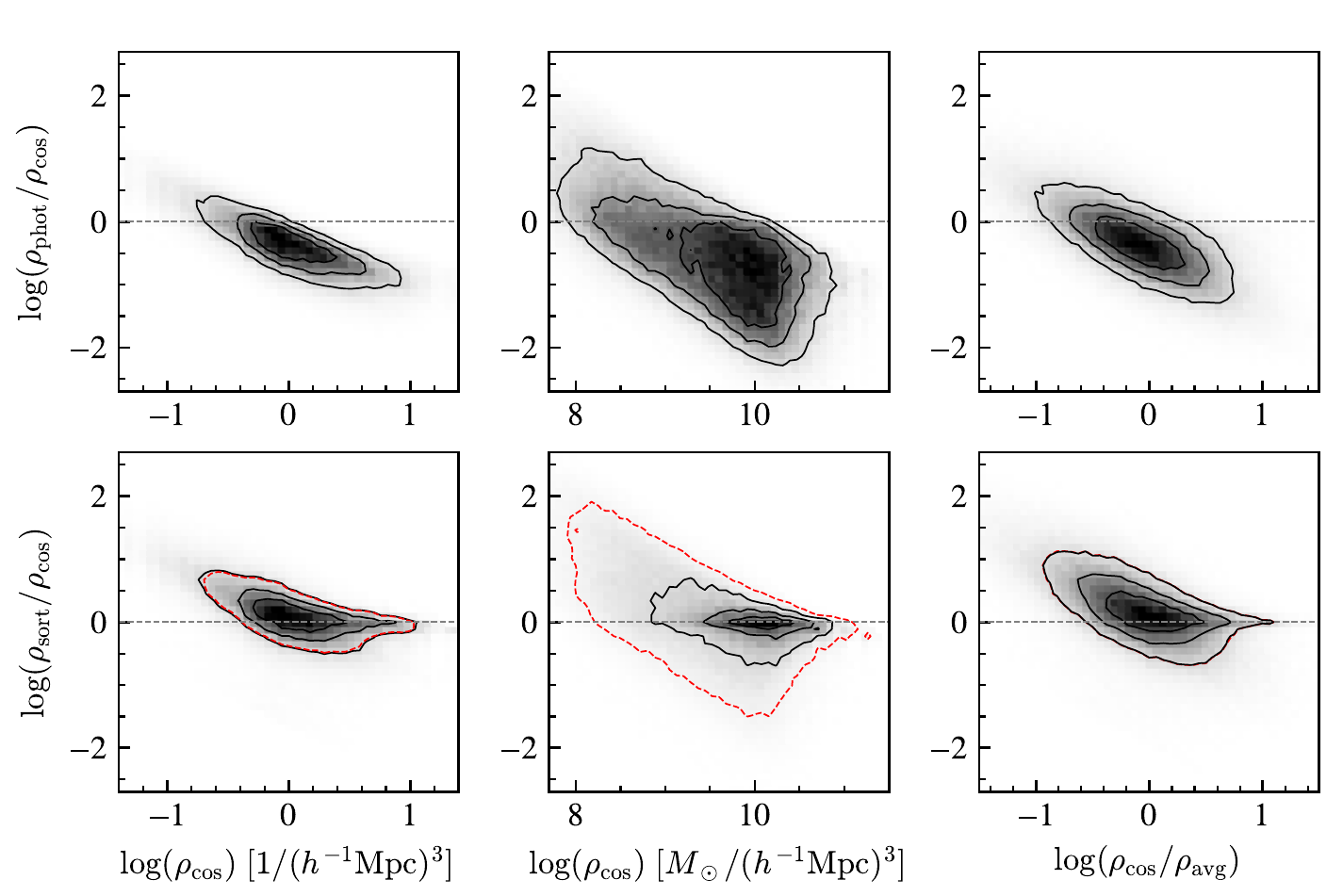}
    \caption{Two-dimensional density error histograms for $\phot{\rho}$ and $\sort{\rho}$. Densities were calculated in cylinders of length 4~$h^{-1}$Mpc. The left panels show number density, the middle panels show stellar mass densities, and the right panels show halo mass densities using only central galaxies. The solid contours represent limits of 25, 50, and 75 per cent of the maximum bin value in each subplot. The dashed contour (red) is set at a limit equal to the minimum contour level in the corresponding $\phot{\rho}$ subplot. The horizontal dashed line represents zero error. While \SORT{} struggles with lower densities, we observe much improvement from the highest densities down to average densities. \SORT{} distributions show better alignment with the zero error line while photometric densities all tend to be underestimated except in the lowest-density environments.}
    \label{fig:delta_density_4Mpc}
\end{figure*}

\begin{figure*}
    \centering
    \includegraphics[width=5.6in]{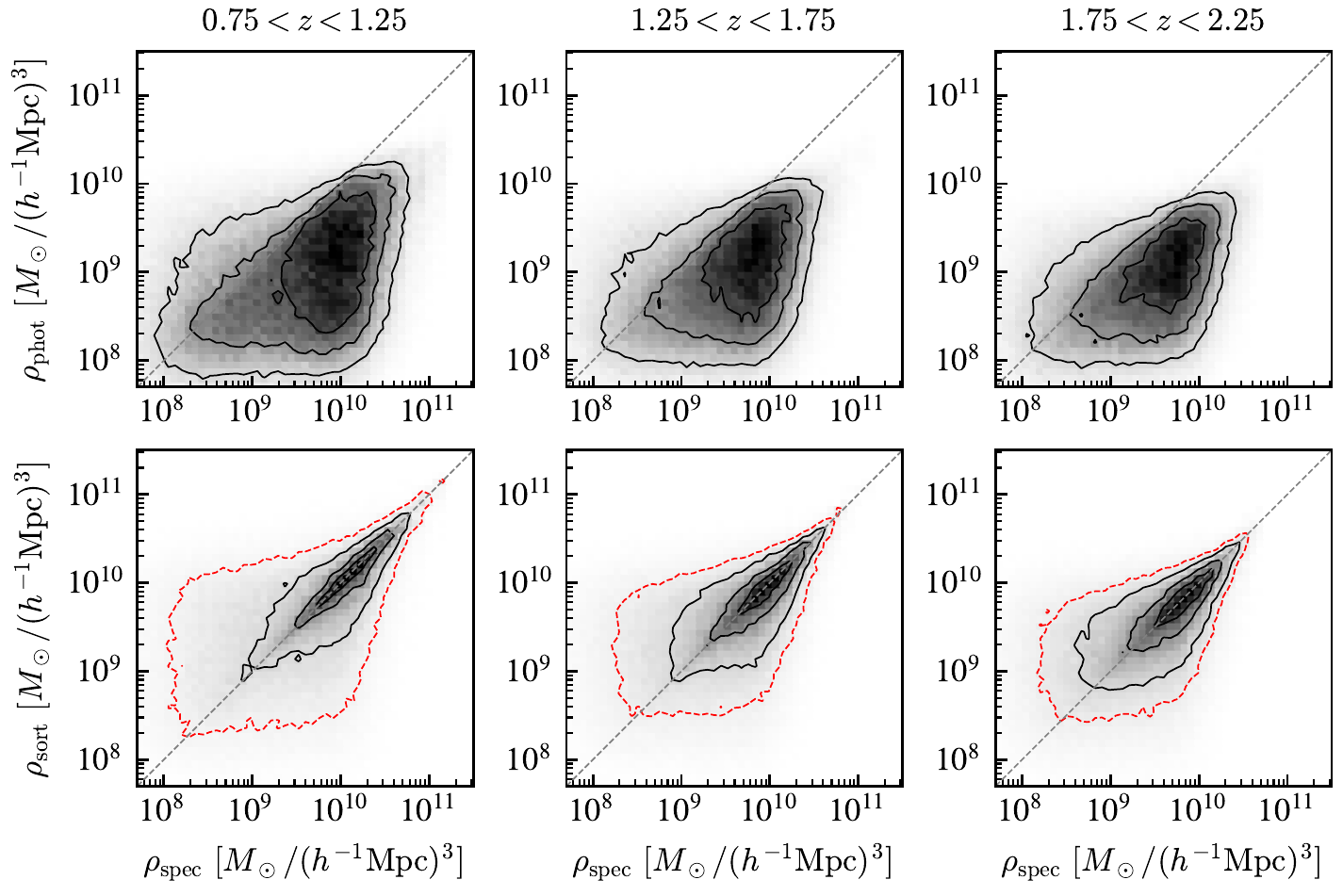}
    \caption{Two-dimensional stellar mass density histograms for $\phot{\rho}$ (top panels) and $\sort{\rho}$ (bottom panels) for all redshift ranges of the light cone using a cylinder length of $4~h^{-1}$Mpc. The solid contours represent limits of 25, 50, and 75 per cent of the maximum bin value in each subplot. The dashed contour (red) is set at a limit equal to the minimum contour level in the corresponding $\phot{\rho}$ subplot. We observe consistent improvement in density estimates with \SORT{} at all redshifts. Biases in regions of average or higher density are greatly reduced. \SORT{} distributions are more symmetric across the line of equality and overall scatter is lower.}
    \label{fig:stellar_mass_density}
\end{figure*}

\begin{figure*}
    \centering
    \includegraphics[width=5.6in]{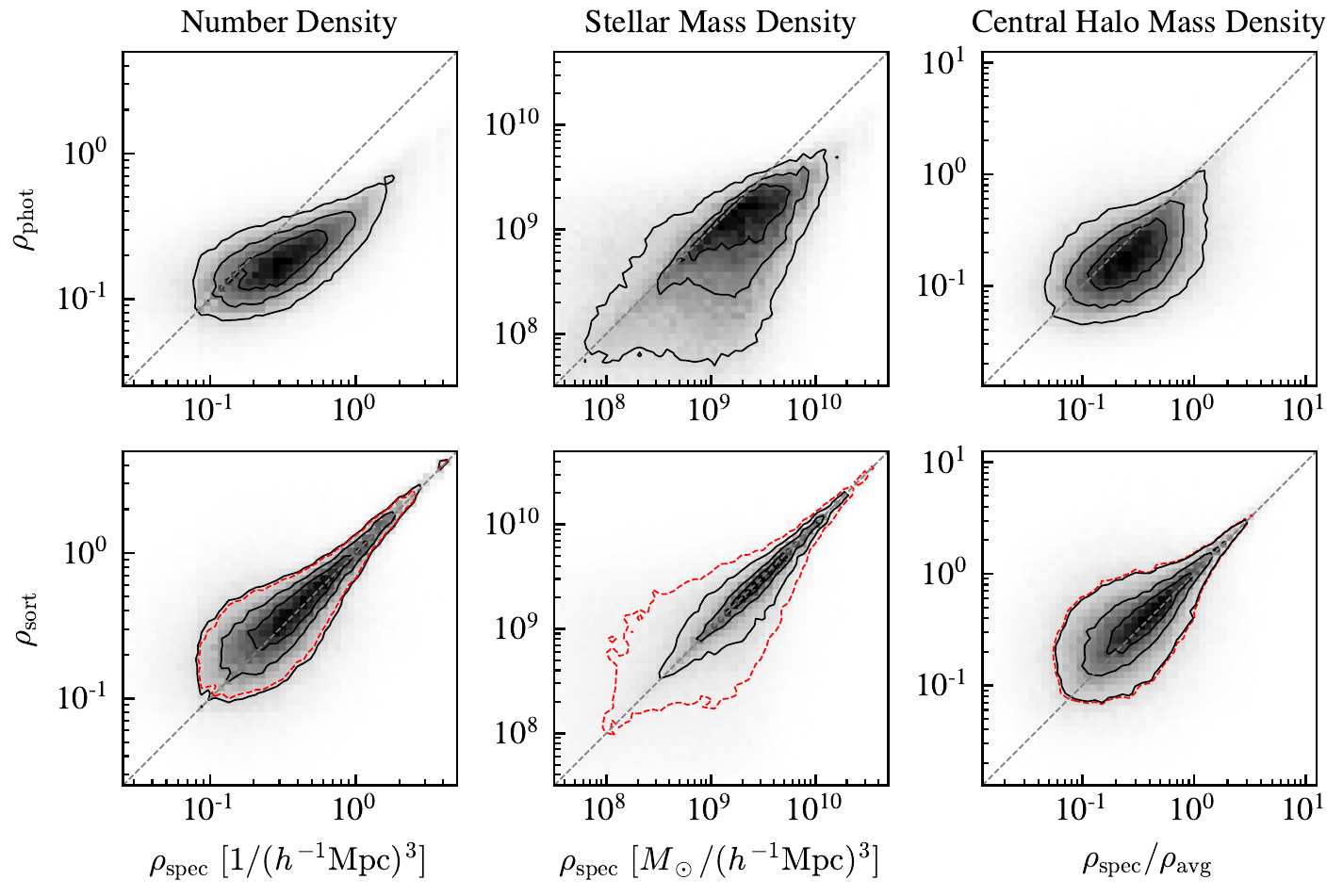}
    \caption{Two-dimensional density histograms for $\phot{\rho}$ (top panels) and $\sort{\rho}$ (bottom panels) in the range $0.75<z<1.25$. Densities were calculated within cylinders of length $l=2\frac{1000~\text{km~s}^{-1}}{c}(1+z)$. The left panels show number density, the middle panels show stellar mass densities, and the right panels show halo mass densities using only central galaxies. The solid contours represent limits of 25, 50, and 75 per cent of the maximum bin value in each subplot. The dashed contour (red) is set at a limit equal to the minimum contour level in the corresponding $\phot{\rho}$ subplot. The longer length of the cylinder significantly improves \SORT{} density estimates, most notably for stellar mass densities.}
    \label{fig:all_density_1000kms}
\end{figure*}

\begin{figure*}
    \centering
    \includegraphics[width=0.75\linewidth]{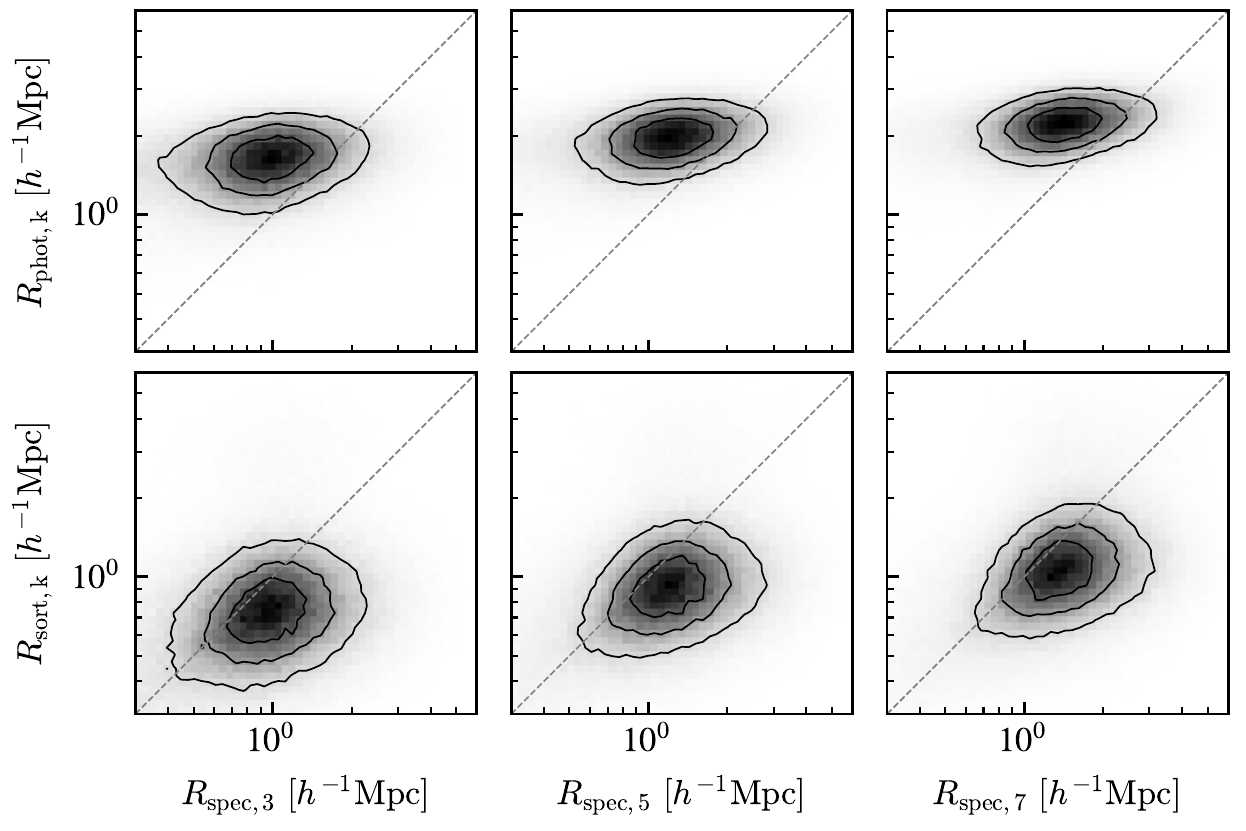}
    \caption{Two-dimensional histograms of 3D distances to $k$th nearest neighbors using $\phot{z}$ (top panels) and $\sort{z}$ (bottom panels) compared to $\spec{z}$ for $k=3$, 5, and 7. Overall scatter is slightly increased using $\sort{z}$, but alignment with the line of equality is improved, particularly at smaller scales. At larger scales (corresponding to lower densities), \SORT{} underestimates $R_k$ as it packs galaxies too closely together in low-density environments. See Section~\ref{section:future_considerations} for details.}
    \label{fig:r_nn}
\end{figure*}

Here we provide additional supplementary figures that support the main text. Fig.~\ref{fig:z_diff} shows two-dimensional histograms of the redshift errors as a function of the chosen redshift -- either $\phot{z}$ or $\sort{z}$. Fig.~\ref{fig:dz_hist_fg} shows redshift errors for \SORT{} with varying spectroscopic fractions. Fig.~\ref{fig:2pcf_full_range} shows results of the \tpcf{} in each of the three complete redshift bins. Fig.~\ref{fig:square_field_alt} shows another square region of space (similar to Fig.~\ref{fig:square_field}) using different redshift types. Fig.~\ref{fig:individual_z_dist2d} shows two-dimensional redshift histograms for each of the three complete redshift bins. Fig.~\ref{fig:delta_density_4Mpc} shows two-dimensional histograms that correspond to the errors of densities shown in Fig.~\ref{fig:all_density_4Mpc}. Fig.~\ref{fig:stellar_mass_density} shows stellar mass densities in each of the three complete redshift bins using cylinders of length $4~h^{-1}$Mpc. Fig.~\ref{fig:all_density_1000kms} shows densities similar to Fig.~\ref{fig:all_density_4Mpc} but calculated at a larger length scale of $l=2\frac{1000 \text{km s}^{-1}}{c}(1+z)$. Fig~\ref{fig:r_nn} show two-dimensional histograms of 3D distances to $k$th nearest neighbors for $k=3$, 5, and 7.

\bsp	
\label{lastpage}
\end{document}